\PassOptionsToPackage{unicode}{hyperref}
\PassOptionsToPackage{hyphens}{url}
\documentclass[
]{article}
\usepackage{amsmath,amssymb}
\usepackage{iftex}
\ifPDFTeX
  \usepackage[T1]{fontenc}
  \usepackage[utf8]{inputenc}
  \usepackage{textcomp} 
\else 
  \usepackage{unicode-math} 
  \defaultfontfeatures{Scale=MatchLowercase}
  \defaultfontfeatures[\rmfamily]{Ligatures=TeX,Scale=1}
\fi
\usepackage{lmodern}
\ifPDFTeX\else
\fi
\IfFileExists{upquote.sty}{\usepackage{upquote}}{}
\IfFileExists{microtype.sty}{
  \usepackage[]{microtype}
  \UseMicrotypeSet[protrusion]{basicmath} 
}{}
\makeatletter
\@ifundefined{KOMAClassName}{
  \IfFileExists{parskip.sty}{%
    \usepackage{parskip}
  }{
    \setlength{\parindent}{0pt}
    \setlength{\parskip}{6pt plus 2pt minus 1pt}}
}{
  \KOMAoptions{parskip=half}}
\makeatother
\usepackage{xcolor}
\usepackage[margin=1in]{geometry}
\usepackage{longtable,booktabs,array}
\usepackage{calc} 
\usepackage{etoolbox}
\makeatletter
\patchcmd\longtable{\par}{\if@noskipsec\mbox{}\fi\par}{}{}
\makeatother
\IfFileExists{footnotehyper.sty}{\usepackage{footnotehyper}}{\usepackage{footnote}}
\makesavenoteenv{longtable}
\usepackage{graphicx}
\makeatletter
\def\maxwidth{\ifdim\Gin@nat@width>\linewidth\linewidth\else\Gin@nat@width\fi}
\def\maxheight{\ifdim\Gin@nat@height>\textheight\textheight\else\Gin@nat@height\fi}
\makeatother
\setkeys{Gin}{width=\maxwidth,height=\maxheight,keepaspectratio}
\makeatletter
\def\fps@figure{htbp}
\makeatother
\providecommand{\tightlist}{%
  \setlength{\itemsep}{0pt}\setlength{\parskip}{0pt}}
\NewDocumentCommand\citeproctext{}{}
\NewDocumentCommand\citeproc{mm}{%
  \begingroup\def\citeproctext{#2}\cite{#1}\endgroup}
\makeatletter
 \let\@cite@ofmt\@firstofone
 \def\@biblabel#1{}
 \def\@cite#1#2{{#1\if@tempswa , #2\fi}}
\makeatother
\newlength{\cslhangindent}
\newlength{\csllabelwidth}
\newenvironment{CSLReferences}[2] 
 {\begin{list}{}{%
  \setlength{\itemindent}{0pt}
  \setlength{\leftmargin}{0pt}
  \setlength{\parsep}{0pt}
  \ifodd #1
   \setlength{\leftmargin}{\cslhangindent}
   \setlength{\itemindent}{-1\cslhangindent}
  \fi
  \setlength{\itemsep}{#2\baselineskip}}}
 {\end{list}}
\usepackage{calc}

\usepackage{setspace}
\usepackage{booktabs}
\usepackage{pdflscape}
\newcommand{\blandscape}{\begin{landscape}}
\newcommand{\elandscape}{\end{landscape}}
\usepackage{booktabs}
\usepackage{longtable}
\usepackage{array}
\usepackage{multirow}
\usepackage{wrapfig}
\usepackage{float}
\usepackage{colortbl}
\usepackage{pdflscape}
\usepackage{tabu}
\usepackage{threeparttable}
\usepackage{threeparttablex}
\usepackage[normalem]{ulem}
\usepackage{makecell}
\usepackage{xcolor}
\ifLuaTeX
  \usepackage{selnolig}  
\fi
\usepackage{bookmark}
\IfFileExists{xurl.sty}{\usepackage{xurl}}{} 
\hypersetup{
  pdftitle={Data inaccuracy quantification and uncertainty propagation for bibliometric indicators},
  pdfauthor={Paul Donner},
  hidelinks,
  pdfcreator={LaTeX via pandoc}}

\title{Data inaccuracy quantification and uncertainty propagation for bibliometric indicators}
\author{Paul Donner}
\date{June 2024}

\begin{document}
\maketitle

\textbf{Abstract}

This study introduces an approach to estimate the uncertainty in bibliometric indicator values that is caused by data errors. This approach utilizes Bayesian regression models, estimated from empirical data samples, which are used to predict error-free data. Through direct Monte Carlo simulation -- drawing many replicates of predicted data from the estimated regression models for the same input data -- probability distributions for indicator values can be obtained, which provide the information on their uncertainty due to data errors. It is demonstrated how uncertainty in base quantities, such as the number of publications of a unit of certain document types and the number of citations of a publication, can be propagated along a measurement model into final indicator values. Synthetic examples are used to illustrate the method and real bibliometric research evaluation data is used to show its application in practice. Though in this contribution we just use two out of a larger number of known bibliometric error categories and therefore can account for only some part of the total uncertainty due to inaccuracies, the latter example reveals that average values of citation impact scores of publications of research groups need to be used very cautiously as they often have large margins of error resulting from data inaccuracies.


\section{Introduction}\label{introduction}

The formal analysis of uncertainty in results due to data errors is rarely considered in quantitative science studies and evaluative bibliometrics. We know only very little about how accurate and robust scientometric measurements and results are in light of pervasive data errors. Knowing the measurement quality can lead to an improved understanding of the fundamentals of the science system and its dynamics: when we become aware of the measurement error of instruments and methods, we can better appreciate their affordances and limitations, and therefore the limitations of the results that were obtained with them. A crucial aspect of measurement quality in scientometrics is the quality of the original data. Sufficient data quality of third-party data is routinely taken for granted and rarely scrutinized independently, even though a considerable number of studies have brought to light severe and persistent levels of error in basic bibliometric data. Known data quality issues are usually not integrated into research, for example by sensitivity analysis or reporting error margins for result values. The bibliometrics and research evaluation community is aware of this deficiency to some degree. For instance, the influential Leiden Manifesto for research metrics (\citeproc{ref-hicks2015LM}{Hicks et al. 2015}), in principle 8, establishes the following straightforward precept: `If uncertainty and error can be quantified, for instance using error bars, this information should accompany published indicator values. If this is not possible, indicator producers should at least avoid false precision.' Nonetheless, it is exceedingly rare to come across bibliometric studies which include quantification of uncertainty due to data errors. In part, this is likely because of a lack of an accepted method to calculate and express these uncertainties.

In contrast, it is a widely accepted convention of the physical sciences that the result of a measurement consists of the best estimate of the value of a quantity and the uncertainty of that estimate (\citeproc{ref-taylor1997introduction}{Taylor 1997}), for example, as in the relative uncertainties of measured fundamental physical constants in Tiesinga et al. (\citeproc{ref-tiesinga2021codata}{2021}). Reported uncertainties are quantified statements on the quality of a measurement and allow users to judge its reliability. Known uncertainties of input quantities can be propagated along measurement models to obtain the uncertainties of output quantities by well understood procedures, called uncertainty propagation (\citeproc{ref-jcgm2008evaluation}{JCGM 2008}; \citeproc{ref-taylor1997introduction}{Taylor 1997}).

It is possible to apply this framework of uncertainty and its propagation to the social sciences, even though they are not dealing with measurement in the physical sense. As an example for translating this concept to scientometrics, consider the following. The citation counts of individual publications recorded in bibliometric databases are not always accurate. These errors can be modelled and described by a probability distribution. The uncertainties of the citation counts of a number of individual publications and a specified indicator formula determine the uncertainty of the value of that citation impact indicator, such as the average number of citations of the publications of a research organization. The indicator formula or algorithm corresponds to the measurement model of physical science uncertainty propagation. If the distribution of the citation count error is known and can be formalized in a stochastic model, its influence on indicator results can be quantified by simulation studies using the Monte Carlo method. How is this useful? For instance, if the difference of the measured citation impact scores of two compared units of assessments is on the order of 10 per cent, it is important to know if the uncertainties of the two compared indicator result values are on the order of 2 per cent or 20 per cent. Only in the first case can an interpretation of the observed difference in the scores be justified as reliable. Uncertainty estimates based on data error distributions permit a judgment on whether the available data is accurate enough to achieve valid, robust, and useful results. This is the rationale for studying uncertainty propagation for bibliometric data and indicators.

The approach to uncertainty propagation for bibliometric indicator values proposed in this paper can be summarized abstractly by the following sequence of steps.

\begin{enumerate}
\def\labelenumi{\arabic{enumi}.}
\tightlist
\item
  Define the types of error or data quality issues whose influence on bibliometric scores is to be quantified.
\item
  Collect an appropriate empirical sample. Determine the correct values of the variables in question in addition to the observed, possibly erroneous values.
\item
  Specify a Bayesian regression model structured to predict the error-free values of the variable from the observed values and any relevant available covariates. Estimate the model with the collected empirical data.
\item
  Use the estimated model for Bayesian posterior prediction of the likely error-free values based on the observed values of the variable in new data on the micro level of individual items a large number of times (Monte Carlo simulation).
\item
  Calculate bibliometric scores of interest from the predicted values for each Monte Carlo iteration.
\item
  Summarize and report the obtained indicator score distributions.
\end{enumerate}

This approach is exemplified in the following sections for two sources of error, omitted citations and errors in document type assignments. As discussed below, besides these there are many more error sources which are not included in this study, which means the expressions of uncertainty reported here only reflect a certain part of the total uncertainty arising from data inaccuracies. Such total uncertainties integrating all known such sources of data errors would realistically be of greater magnitudes, thus this study's uncertainty quantifications should be understood as partial and likely underestimating the full uncertainties.

The rest of the paper is structured as follows. In the next section the literature on uncertainty in scientometrics and on errors in bibliometric data is reviewed. Following that, we present empirical sample data for omitted citations. This data is then used in section \ref{prop} to estimate Bayesian regression models of the error distribution of citation counts. This section also introduces the proposed approach in systematic manner. The stochastic error models are used to stage example simulations that illustrate how practitioners and researchers can use this approach to estimate the influence of data quality issues on study results. Conclusions are discussed in the final section.

\section{Related work}\label{related-work}

\subsection{Uncertainty in scientometrics not related to data errors}\label{uncertainty-in-scientometrics-not-related-to-data-errors}

The present object of investigation, the quantification of the uncertainty of indicator values which is due to data errors, should be clearly distinguished from certain other concepts which are concerned with uncertainty which play a role in scientometrics.

In cases when the investigated objects form a well-defined population and it is impractical or not feasible to analyze all objects one can resort to statistical sampling. When a random sample is studied, the values of quantities calculated on this subset of the population do not precisely match the values which could be calculated for the same quantity for the whole population, which is the actual quantity of interest. It is therefore important to quantify how similar the sample and population values are expected to be, in other words, the uncertainty of the sample value in representing the population value, i.e.~the statistical sampling error. This uncertainty depends on the specific sampling method and sample size. Sampling uncertainty can be quantified, for example, by frequentist statistical confidence intervals for estimates. Some examples of scientometric applications of this approach are Schubert \& Glänzel (\citeproc{ref-schubert1983statistical}{1983}) and Schubert et al. (\citeproc{ref-schubert1988against}{1988}). Sampling uncertainty conveys important information on how well a random subset can be expected to represent a population based on insights from probability theory. It is not concerned in any way with errors in the values of data.

Waltman et al. (\citeproc{ref-waltman2012leiden}{2012}) introduced stability intervals for values of bibliometric indicators. These intervals are intended to quantify the relative susceptibility of an indicator value to minor changes in the underlying publication data, in other words, they reflect how stable or volatile the actually observed value is with respect to the observed data. If a bibliometric indicator value is highly dependent on relatively few input values, its stability is considered low and its stability interval is relatively large. Stability intervals are operationalized by means of bootstrapping, that is, drawing repeated samples with replacement from the available data, where the resampled number of observations is equal to the original number of observations, and computing the statistic of interest for each sample. The width of the bootstrap intervals of statistics can be computed in various ways from the distribution of resampled values of the statistic of interest (replicates), most simply by using the 0.025- and 0.975-quantiles of the bootstrap distribution to specify the 95 per cent bootstrap confidence interval, which is the method preferred by Waltman et al. (\citeproc{ref-waltman2012leiden}{2012}).

This approach has some conceptual difficulties. Because the bootstrap method is intended for statistical inference from random samples of larger populations (\citeproc{ref-efron1994introduction}{Efron \& Tibshirani 1993 p. 17}), it fundamentally is based on the assumption that the set of observations could have been different. Bootstrapping, being a non-parametric method, frees one of the necessity to assume a specific distribution of the values in a sample and can be used to do inference for problems for which no analytical solution is available. Yet it does not do away with the prerequisite of sampling. This assumption of random sampling from a larger population is generally not met in bibliometric applications, which is one of the reasons why null hypothesis significance testing in research evaluation was criticized by Schneider (\citeproc{ref-schneider2013caveats}{2013}). When the stability interval is calculated for a bibliometric indicator summarizing the citation distribution of a research institution or journal, the entire population of publications can usually be observed or otherwise the observed subset is not a random sample and it is not clear how the set of observations could have been different -- certainly not by drawing another sample from the same population. Moreover, resampling with replacement on such a data set suggests that publications could have not been produced or could have been produced more than once while they were in fact produced exactly once.

The stability of bibliometric indicator values which are averages is related to the number of observations -- more observations lead to more stable values -- and the skewness of the distribution -- more skew leads to more volatile values. Notwithstanding the above conceptual reservations, stability intervals have been shown to be of practical utility as they do in fact show quite clearly those values which are greatly affected by few unusual observations as the examples in Waltman et al. (\citeproc{ref-waltman2012leiden}{2012}) demonstrate.

\subsection{Accuracy and completeness of citation links in bibliographic databases}\label{accuracy-and-completeness-of-citation-links-in-bibliographic-databases}

Bibliometric data from commercial data sources such as Web of Science (\citeproc{ref-birkle2020web}{Birkle et al. 2020}) or Scopus (\citeproc{ref-baas2020scopus}{Baas et al. 2020}) and non-commercial sources like OpenAlex (\citeproc{ref-priem2022openalex}{Priem et al. 2022}) is widely used in both academic science studies and in applied research assessment, monitoring, and funding distribution. An example is the Science \& Engineering Indicators report series of the U.S. National Science Foundation's National Science Board, which has used Science Citation Index data for many years and switched to Scopus data in 2016. Similar monitoring reports are produced with bibliometric data in other countries (\citeproc{ref-moed2005citation}{Moed 2005 p. 271}). Performance-based funding systems in many countries include a bibliometric component (\citeproc{ref-hicks2012performance}{Hicks 2012}; \citeproc{ref-zacharewicz2019performance}{Zacharewicz et al. 2019}). Various global university rankings base their scores on bibliometric data. For instance, the Times Higher Education and the QS rankings currently use Scopus data while ShanghaiRanking and U.S. News \& World Report use WoS data.

Bibliometric data sources, including commercially produced ones, are not error-free (e.g. \citeproc{ref-pranckute2021web}{Pranckutė 2021 pp. 20--4}). There are a number of types of errors or inaccuracies in bibliometric data that can exert adverse influence on the results of bibliometric studies. In this paper, only two types are studied: incomplete citation counts and incorrect document type assignments. Other types of errors include, but are not limited to

\begin{itemize}
\tightlist
\item
  incompletely or incorrectly identified authors and institutions (\citeproc{ref-dickersin2002problems}{Dickersin et al. 2002}; \citeproc{ref-donner2020comparing}{Donner et al. 2020}; \citeproc{ref-schulz2016using}{Schulz 2016}),
\item
  incorrect or missing publication identifiers (\citeproc{ref-franceschini2015errors}{Franceschini et al. 2015}),
\item
  duplicate records (\citeproc{ref-liu2021same}{Liu et al. 2021}; \citeproc{ref-valderrama2015systematic}{Valderrama-Zurián et al. 2015}) and missing records (\citeproc{ref-krauskopf2019missing}{Krauskopf 2019}),
\item
  insufficiently standardized publication venue data (\citeproc{ref-leydesdorff2016professional}{Leydesdorff et al. 2016 pp. 2134--6}),
\item
  ignored citation data in supplemental information, a type of missing citations (\citeproc{ref-garcia2015online}{García-Pérez 2015}),
\item
  incomplete or erroneous funding data (\citeproc{ref-alvarez2017funding}{Álvarez-Bornstein et al. 2017}; \citeproc{ref-liu2020accuracy}{Liu 2020})
\end{itemize}

and other missing data elements such as classification assignments of publication records.

Citation counts of scientific research contributions are a fundamental type of data in scientometrics. Accuracy and completeness of citation links are therefore crucial data quality issues (\citeproc{ref-moed2005citation}{Moed 2005}, ch.~13; \citeproc{ref-olensky2016evaluation}{Olensky et al. 2016 pp. 2550--1}). It has been demonstrated that citation counts in bibliometric data sources are subject to non-negligible error rates in a number of studies.

Moed \& Vriens (\citeproc{ref-moed1989possible}{1989}) studied causes of unmatched references to a sample of about 4500 target records from five journals in an online version of the Science Citation Index database. Note that in this study, the reference matching procedure of the Science Citation Index was not studied at all. Rather, it was investigated why for some cited references no exact match could be found even though the cited journal and year of the cited reference were covered in the data. Selecting citing references by cited publication year and cited journal name, including variations, they focused on the data fields cited first author name, volume number, and starting page. They found that about 9 per cent of references contained at least one discrepancy in these fields as compared to the identified corresponding target record. Some particular journal conventions, such as `combined volumes' and special pagination for letter sections, could be identified as associated with certain error types, as were some types of author names, particularly those consisting of more than a single word. In auxiliary analyses the authors also established that discrepancies between publication year of the target item and cited reference are common. The results demonstrate convincingly that any strict reference matching using only exact agreement of field values would overlook a considerable number of citation links and that therefore reference matching algorithms in general need to be lenient with respect to minor variations in field values between cited and citing record in order to attain high completeness of citation counts. Moreover, there are clear systematics at play both for entire target journals and for individual publications, which precludes any assumption that these errors can be neglected because they are distributed randomly and affect all cited documents equally.

Another early empirical study is that of Abt (\citeproc{ref-abt1992fraction}{1992}) who looked up around 1000 cited references from one astrophysical journal issue to eight cited journals using the original volumes or published indexes of the cited literature. Some 12 per cent of cited references were found to be inaccurate in the original reference lists. Abt found that most of these were corrected during processing for Science Citation Index: `the 12.2\% author errors result in only 3.6\% misplacements or omissions in SCI' (p.~236).

Buchanan (\citeproc{ref-buchanan2006accuracy}{2006}) investigated the accuracy of cited references in two citation databases, Science Citation Index Expanded (SCIE) and SciFinder Scholar, using as source data some 6000 cited references from around 200 articles in three chemistry journals. Checking both the original publications' reference lists and the indexed versions allowed him to attribute the cause of found errors to either the authors or to processing by the citation databases. In both databases, around 10 per cent of cited references were not matched to their indexed target. The major reason were author errors but database mapping errors were a considerable factor as well. Among SCIE database mapping errors, the predominant type was omission of cited references. This study confirmed systematic errors for specific journals and individual articles.

Franceschini et al. (\citeproc{ref-franceschini2013novel}{2013}) propose a method to study the error rate of citation links at a large scale with an automated approach by comparing sets of citing and cited papers that are covered by two or more databases, similar to the earlier short study of Dyachenko \& Pislyakov (\citeproc{ref-dyachenko2010scientometric}{2010}). As it is automated, the advantages of the method are that it can be applied directly to any set of publications of interest and that it provides individual paper-level error estimates. Drawbacks are that, first, subscriptions to multiple databases are required and second, that, because the shared citing publications are not independent, it is likely that for cited references with severe inaccuracies, more than a single database would fail to link it to the intended target. If all studied databases cannot resolve a cited reference, the method cannot detect that specific error. Using as a test case articles from three journals from publication year 2008 and the two databases Web of Science (WoS) and Scopus, the study finds error rates of 3.2 per cent and 5.6 per cent, respectively, for this data set. Furthermore they propose a statistical model for estimating the true number of citations from the empirical error rate and the number of observed citations.\footnote{As the equations of Franceschini et al. (\citeproc{ref-franceschini2013novel}{2013}) for the expected value (Eq. 12) and variance (Eq. 13) of the number of omitted citations \(o_{i}\) are multiplicative in \(c_{i}\), the number of observed citations, their model has the limitation that any uncited publication has an expected value of zero omitted citations with zero variance. The same holds for their equations for these variables for sets of papers if the total sum of citations is zero.}

In the first study of its kind, Olensky et al. (\citeproc{ref-olensky2016evaluation}{2016}) studied the performance of two proprietary reference matching algorithms on WoS data compared with the results of the algorithm of WoS itself. Because of the prevalence of inaccuracies in cited reference data and the conservativeness of the WoS algorithm, the two proprietary algorithms exhibited substantially higher Recall at the cost of slightly lower Precision in a sample of about 4000 citations to 300 target papers. While the WoS algorithm missed about 6 per cent of reference matches, the two other algorithms missed about 1 per cent each. These outcomes show that the WoS algorithm is apt to make no or very few false positive errors but overlooks inaccurate references leading to false negatives, which can be overcome with alternative matching strategies. The study also found a number of citation links present in the data which were not based on cited references in the original publications.

Van Eck \& Waltman (\citeproc{ref-vaneck2017accuracy}{2017}) studied the representation of a large data set of cited references of papers published in journals by the publishing house Elsevier in the citation databases WoS and Scopus. One important finding is that in WoS the number of cited references is lower than that of the original publication for 19 per cent of matched papers while the corresponding figure for Scopus is only 1 per cent. Both databases had a similar share of papers without indexed references (about 1 per cent) which do have cited references in the original publications. Inspection of a sample showed that both databases are prone to miss cited references present in the original data.

Further studies concerned with cited reference data errors include Tunger et al. (\citeproc{ref-tunger2010delphic}{2010}), Schmidt (\citeproc{ref-schmidt2018fehler}{2018}), Dyachenko \& Pislyakov (\citeproc{ref-dyachenko2010scientometric}{2010}), Olensky (\citeproc{ref-olensky2015data}{2015}), and Tüür-Fröhlich (\citeproc{ref-tuur2016non}{2016}).

While in the past decades citation matching had to rely on bibliographic metadata, such as author names, source and publication titles, publication year and volume number, now most publications are published with unique unambiguous identifiers, Digital Object Identifiers (DOIs). DOIs have also been assigned retrospectively to a large share of published scientific literature and they are widely accepted as part of reference lists. As they in principle permit an error-free matching of cited a reference to a record of an indexed document, one might expect that the error in citation matching is much reduced for the more recently published literature for which citation database vendors can utilize DOI matching. However, the data of identifiers is just as much subject to data quality problems as any other data (\citeproc{ref-franceschini2015errors}{Franceschini et al. 2015}, \citeproc{ref-franceschini2016museum}{2016}; \citeproc{ref-krauskopf2023inconsistency}{Krauskopf \& Salgado 2023}; \citeproc{ref-zhu2019doi}{Zhu et al. 2019}). For instance, Huang et al. (\citeproc{ref-huang2020comparison}{2020}) compared three bibliometric data sources and found some DOIs only present in one of them. Checking a sample of them manually, they report:

\begin{quote}
We have also checked each DOI against the DOI string actually recorded as per original document (where applicable) or via doi.org. These percentages (of correct DOIs) are 93.1\%, 98.2\% and 96.7 for WoS, Scopus, and MSA respectively (with all 15 institutions combined). While these numbers are relatively high, a significant number of errors suggests that DOIs are not being systematically checked against authoritative sources such as Crossref which we find surprising.
\end{quote}

DOI errors have also been found to be prevalent in the cited references data (\citeproc{ref-cioffi2022identifying}{Cioffi et al. 2022}; \citeproc{ref-xu2019types}{Xu et al. 2019}), confirming that there data quality issues on both the citing and the cited sides of the reference matching task. Clearly, the introduction and widespread use of unique identifiers in cited reference data has not solved the citation matching problem.

From this cross-section overview of the literature the following conclusions can be drawn:

\begin{enumerate}
\def\labelenumi{\arabic{enumi}.}
\tightlist
\item
  References in the original publications are often incorrect (\citeproc{ref-dubin2004most}{Dubin 2004}; \citeproc{ref-jergas2015quotation}{Jergas \& Baethge 2015}; \citeproc{ref-moed1989possible}{Moed \& Vriens 1989}; \citeproc{ref-sauvayre2022misreferencing}{Sauvayre 2022}; \citeproc{ref-wager2008technical}{Wager \& Middleton 2008}).
\item
  Errors in citation matching in citation indexes can have several causes. Besides the above item 1, a reference in the citation index can be an inaccurate copy of an original correct reference. The reference matching procedure may be unable to resolve some surface form of a reference to another, that is, link a source citation to its intended target.
\item
  The proprietary reference matching procedure of WoS is based on close agreement of cited reference data (normalized and `re-written' during processing (\citeproc{ref-garfield1983quality}{Garfield 1983})) to the target papers' bibliographical data, which is also normalized during source indexing. This corrects a considerable share of errors in cited references data made by citing authors (\citeproc{ref-abt1992fraction}{Abt 1992}; \citeproc{ref-buchanan2006accuracy}{Buchanan 2006}). Consequently, the reference matching procedure of WoS has near-optimal Precision but incomplete Recall -- it is known to miss some inaccurate reference links (\citeproc{ref-olensky2015data}{Olensky 2015}).
\item
  Advanced reference matching algorithms that are more error-tolerant, can identify more correct citation links in the case of WoS data as compared to WoS's own algorithm.
\item
  The share of unmatched references that should ideally be matched is roughly around 5 per cent.
\item
  The indexing process of references by citation databases is susceptible to the introduction of errors such as overlooking references or wrongly correcting erroneous references. The scale of these inaccuracies is more difficult to determine but probably lower than the unmatched reference rate.
\end{enumerate}

Despite these known flaws of reference matching algorithms and inaccurate references in the original publications, usually no attempts are made to incorporate this uncertainty about citation counts into the reporting of indicator results. In the second part of this paper we sketch a method of how this could be done.

What is also missing is an estimate the rate of missed citations by a principled method for a random sample -- in the sense of being representative for a complete citation index database. For the first part of this study, a simple random sample of WoS source papers was drawn and it was attempted to find all reference strings of WoS indexed documents that reference them. Considerable effort was expended on identifying inexact matches. The objective is to give a statistical estimate of the proportion of missed citations and to describe the relationship of the number of found citations to the number of missed citations, i.e.~the conditional error distribution. While we are concerned with the specific case of WoS data in what follows, the problem of suboptimal reference matching is just as acute for other citation index databases.

\section{Missing links: Empirical study of citation error distribution}\label{missing-links-empirical-study-of-citation-error-distribution}

\subsection{Methods and data}\label{methods-and-data}

In this section we report on an empirical study\footnote{Initial results from this study were first reported in Donner (\citeproc{ref-donner2016missing}{2016}). The dataset is available at \url{https://doi.org/10.5281/zenodo.13969973}.} of the prevalence of incorrectly matched references in WoS. Incorrectly matched references result in incorrect citation counts. The analysed data originate from licensed raw data in tagged data format of the WoS journal and proceedings citation indexes\footnote{Science Citation Index Expanded, Social Sciences Citation Index, Arts \& Humanities Citation Index, Conference Proceedings Citation Index - Science, Conference Proceedings Citation Index - Social Sciences \& Humanities. All licensed for records of publication years 1980 to present.} as of spring 2015. A simple random sample of target items published between 1980 and 2015 was drawn from all journal articles, letters and reviews, as defined in the data. Hence, the specific results of this study apply to WoS, not other citation databases. They also do not apply to specific subsets of publications whose error structure is known or presumed to be different, such as publications with references in footnotes, which are more difficult to extract. For such cases, as for other databases, new empirical samples should be collected and analyzed.

Cited references data of all publications from 1980 to spring 2015 were indexed for search. The reference strings, consisting of author name field, split into last name and initials at the comma, the source title, publication year and first page fields were indexed with Oracle Text in an Oracle Database instance. The volume field was not considered because, for the target journal items, volume and publication year are nearly redundant information but publication year is more accurate and more complete in WoS reference data than volume (\citeproc{ref-moed2005citation}{Moed 2005}, Table 13.1).

A procedure was programmed to search the index for references likely referring to the sampled target items. Because the search field data of the target items has to correspond to the way the reference data is prepared and stored in WoS, the target author name and source title were pre-processed. This entailed the abbreviation of first name to initials, the removal of non-letter characters in the name, and, for the source title, using the WoS abbreviation. Where more than one abbreviated title version for the same journal existed in the data and when an additional author group name was available, all possible combinations of those fields' values were used as search input. The procedure performs a fuzzy search on the index and returns a list of unique candidate reference strings that are sufficiently similar to the target input. The search is deliberately lenient so that all possible matches are returned in order to prevent false negatives as much as possible, which is a requirement for this study. This procedure searched for simultaneous fuzzy hits on first author lastname, first author initial (target item co-authors were not considered), source title, publication year (formatted as string), and firstpage. This strategy found 19132 citation records candidates for 372 sample target items. Each record could stand for one or more citations as only unique records were searched and checked.

The candidates were checked for whether they constitute a match to the target or not by a student assistant. Ambiguous candidates were afterwards assessed by the author. Care was taken to avoid false positive matches by querying the database for any exact matches of the candidate reference strings other than the target item. This entire procedure was completely independent of the reference matching done by WoS. The found positive matches are used as citation links and the derived independent citation count for each target item is calculated by retrieving all references using those unique candidate strings.

The WoS-provided citation links were obtained for comparison. They are constituted by the WoS matchkey, designated as the T9/R9 fields in the tagged data files. All citations until 2015 were counted. Table \ref{tab:example} below shows one specific example, indicating which variants were found by the above method for this study and which were available in the original WoS data. Column ``Identity Hits'' indicates whether there is a source record in the database that matches exactly with the reference in all fields. The row ``target source item'' gives the specific form in which a sample publication should appear in the references data according to WoS formatting conventions. Row ``citation 1'' shows that this exact form (combination of family name, initials, year, source, first page) did in fact occur in the data 109 times and was classified as a citation to the target by both the WoS algorithm and our independent inspection. Of the other candidates, the WoS algorithm also counted ``citation 4'', a variant in which the publication year is incorrect and a variant of the source name abbreviation is used. This variant occurred twice. In total WoS found 111 citations for the target item. Besides ``citation 4'', we were also able to identify ``citation 2'' (year error), ``citation 5'' (first page error) and ``citation 7'' (missing first page, variant source abbreviation) as valid citations, which added up to four additional citations. We judged ``citation 3'' as not valid (different year and first page) and ``citation 6'' as not valid (different author name initial, first page and volume; and matches another source record in the database exactly).

\begin{landscape}\begin{table}

\caption{\label{tab:example}Example of reference candidate records}
\centering
\begin{tabular}[t]{>{}l|llll>{}l|llll}
\toprule
Record & Family Name & Initials & Year & Source & First Page & Occurrence & Identity Hit & Match WoS & Match this study\\
\midrule
target source item & SMITH & DS & 1981 & BRIT MED J & 517 &  &  &  & \\
\midrule
citation 1 & SMITH & DS & 1981 & BRIT MED J & 517 & 109 & yes & yes & yes\\
citation 2 & SMITH & DS & 1986 & BRIT MED J & 517 & 1 & no & no & yes\\
citation 3 & SMITH & DS & 1969 & BMJ-BRIT MED J & 90 & 1 & no & no & no\\
citation 4 & SMITH & DS & 1986 & BR MED J & 517 & 2 & no & yes & yes\\
citation 5 & SMITH & DS & 1981 & BRIT MED J & 5 & 4 & no & no & yes\\
citation 6 & SMITH & R & 1986 & BMJ-BRIT MED J & 1043 & 1 & yes & no & no\\
citation 7 & SMITH & DS & 1981 & BR MED J &  & 1 & no & no & yes\\
\bottomrule
\end{tabular}
\end{table}
\end{landscape}

\subsection{Results}\label{results}

\begin{figure}
\centering
\includegraphics{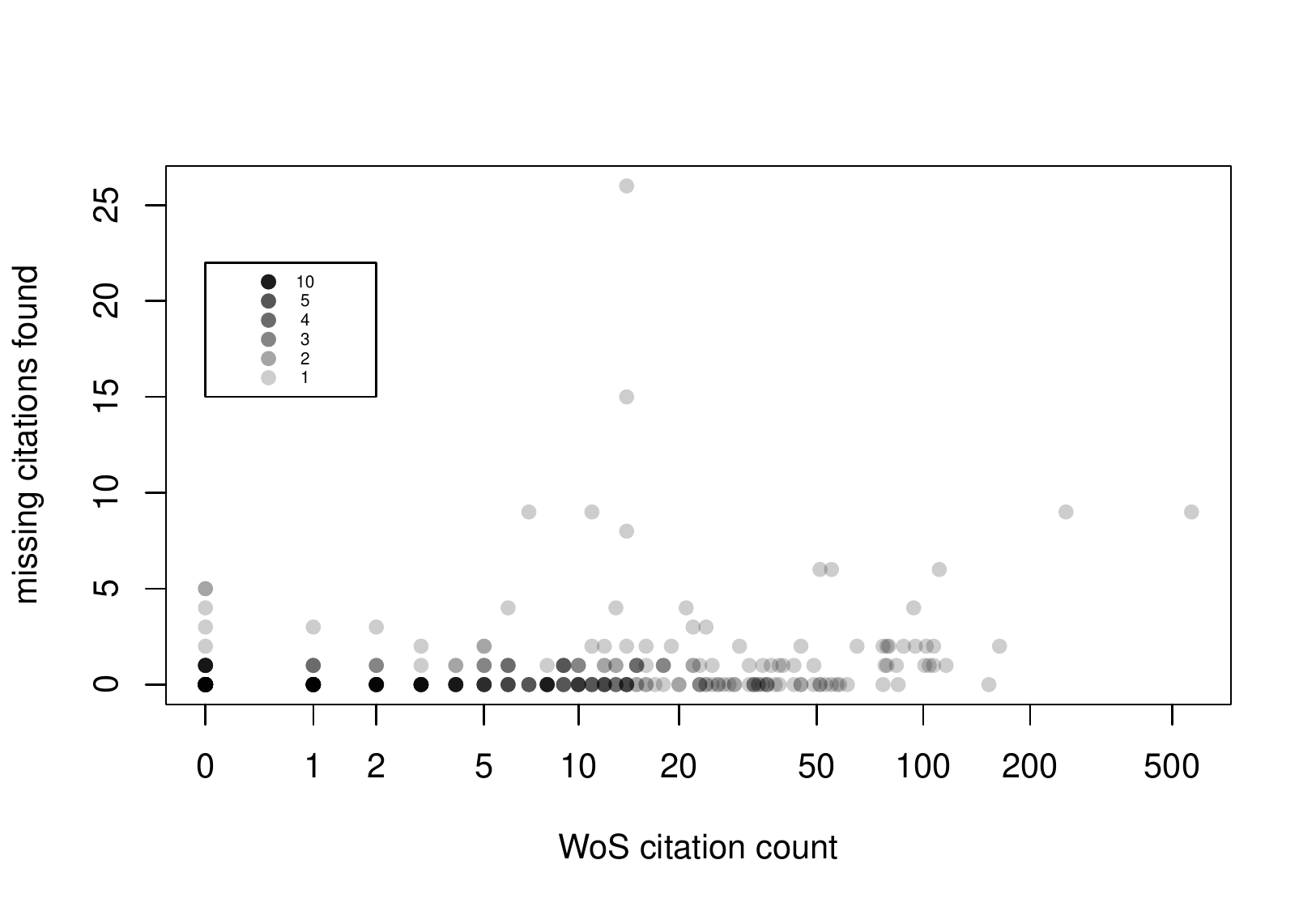}
\caption{\label{fig:sample-plot}Scatterplot of WoS citation count and number of missing citations found}
\end{figure}

A total of 372 sampled WoS records were assessed. According to WoS, these were cited a total of 6120 times. Our search found 255 additional citations, which means that 4 per cent of the citations were missed by the WoS matching algorithm, which is in agreement with the results of other studies (\citeproc{ref-buchanan2006accuracy}{Buchanan 2006}; \citeproc{ref-garcia2010accuracy}{García-Pérez 2010}; \citeproc{ref-olensky2016evaluation}{Olensky et al. 2016}). The distribution of missed citations is presented in Table \ref{tab:sample-distribution}. It is clearly apparent that the distribution of omitted citations is neither symmetric nor normally distributed. The inclusion of these citations would increase the WoS average citations per paper in the sample from 16.4 citations by 0.7 to 17.1. Of the 372 sample target items, 29.4 per cent have one or more detected missing citation. An association between the apparent citation count and the citation count error can be observed, as the number of WoS citations and the number of missed citation per item are moderately correlated with r~=~0.31, 95 per cent confidence interval: {[}0.22,~0.40{]}, see also Figure \ref{fig:sample-plot}. This means that from the perspective of cited items, missing citations due to unsuccessful matching are not random but structured in the sense that those with higher citation impact (more matched citations) also tend to have more missing citations.

Our study design did not allow us to identify any false positive citations, also called ``phantom citations'', i.e.~citation links present in the database where no corresponding cited reference was present in the original publication. This was not possible because we only used the data present in WoS. As mentioned above, WoS transforms the original data and only the transformed data can be accessed. Phantom citation errors have been found by other studies (\citeproc{ref-dyachenko2010scientometric}{Dyachenko \& Pislyakov 2010}; \citeproc{ref-garcia2010accuracy}{García-Pérez 2010}, \citeproc{ref-garcia2011strange}{2011}; \citeproc{ref-olensky2015data}{Olensky 2015 pp. 114--5}; \citeproc{ref-vaneck2017accuracy}{Van Eck \& Waltman 2017}) and are more rare than missing citations. They do constitute one cause of missing citations because they mis-link an original reference to an incorrect cited source record. Because we only used WoS-internal data we also could not assess the phenomenon of references that are not transferred from the original publication to the database record (\citeproc{ref-buchanan2006accuracy}{Buchanan 2006}). This is another, different kind of missing citations than the one addressed in this paper. In our study we are restricted to indexed but unmatched missing citations. Finally, as we used the cited reference source data and compared it with the publication record source, we did not cover references to preprint or working paper versions of a publication, only to the publisher version.

\begin{table}

\caption{\label{tab:sample-distribution}Distribution of number of missed citations from 372 records in WoS}
\centering
\begin{tabular}[t]{rr}
\toprule
missing citations & frequency\\
\midrule
0 & 263\\
1 & 67\\
2 & 20\\
3 & 5\\
4 & 5\\
\addlinespace
5 & 2\\
6 & 3\\
8 & 1\\
9 & 4\\
15 & 1\\
\addlinespace
26 & 1\\
\bottomrule
\end{tabular}
\end{table}

\section{Propagation of uncertainty from data to bibliometric indicators -- a Bayesian regression approach}\label{prop}

\subsection{Incorporating information about uncertainty into statistical models}\label{incorporating-information-about-uncertainty-into-statistical-models}

In the preceding section we have empirically determined the magnitude of one particular source of error of bibliometric data, missing citations due to incomplete reference matching. Consider that we might also collect other samples for other error sources. Having obtained such empirical data on error distributions, it becomes possible to apply this information to other, out-of-sample data to estimate the effect of data inaccuracies on indicator values of interest for this new data. In this section, we propose a methodology for using empirical data on bibliometric error distributions for obtaining uncertainty intervals for bibliometric indicator values on new data from the same data source with the same properties of data inaccuracies. For this purpose we construct statistical models of the error processes which allow to abstract and generalize from the specific empirical sample data so that one can make stochastic predictions of the likely errors in any new data. Such regression models are estimated once and predictive simulations from the models for other data can be used to specify the uncertainty, not only at item micro-level and the average level across papers, but also at the level of specific indicator values at any aggregation, to quantify how precise indicator values are, given known uncertainty structures in the underlying data. This can be done by using the regression models to make statistical predictions for error-corrected values for new input data, such as the likely error-free citation count predicted from observed error-affected citation count values. Making such simulations a large enough number of times for each input observation gives a probabilistic approximation of the distribution of likely error-free outcome values.

This approach is called the Monte Carlo method. In a Monte Carlo simulation, a process is simulated by repeated random sampling of input data to obtain probability distributions of some outcomes of interest. The random variation is a representation of the uncertainty in the modelled relationships. The shape of the probability distributions of the outcome variables reflect the uncertainty in the simulated results. The simulated outcome values' probability distributions after sampling a sufficiently large number of times can then be statistically described and analysed.

It is possible to construct two types of statistical models from empirical data that can be used to understand the effect of data quality on bibliometric indicators. They differ by the direction of prediction: one can either predict erroneous data from error-free data or the reverse. In the models of the first kind, we start with a given data set without any errors. This could either be real data that was thoroughly checked for quality and corrected by hand, or simulated data. We then use a statistical model of the error process, obtained from other empirical sample data, to predict the degree to which empirically observed levels of data quality issues affect bibliometric indicators calculated from the data. This kind of model is useful for testing hypothetical scenarios, such as ascertaining the error rate in raw data that can still be tolerated if one is interested in comparisons of some indicator scores with a specified accuracy. For example one might create a data set of simulated bibliometric ground truth data and use a model of the first kind to test what range of values would be expected if this data is subjected to the process of indexation in bibliometric databases that results in the observed inaccuracies.

An exposition of the use of the models of the first kind is relegated to the appendix in order to keep the main body of the paper focused on what we consider the more immediately and broadly useful practical applications.

In the models of the second kind, we work with given data from a bibliometric data source for which we know the general relevant error rates from earlier empirical error studies but not the specific inaccuracies in the data at hand. We then use prediction from estimated statistical models to generate a large number of simulated hypothetical error-free versions of the data, which are consistent with the given data and the knowledge about error distributions and calculate hypothetical versions of the relevant bibliometric indicators from the simulated data. The distributions of the simulated indicator values contain the information about the uncertainty of the indicator values which can be summarized, such as with point estimates accompanied by 95 per cent probability credible intervals as error margins of error-free indicator values. That means that these models work in the inverse direction from the models of the first kind. While the models of the first kind start with error-free data and simulate realistic errors, the models of the second kind start with error-affected data and simulate realistic error-free data. Both kinds of models can be estimated from the same empirical data sets, only the dependent and independent variables are switched.

\subsection{A simulation exercise of bibliometric uncertainty propagation -- Simulating error-free data}\label{a-simulation-exercise-of-bibliometric-uncertainty-propagation-simulating-error-free-data}

The typical situation is that researchers have to work with data which is known to be inaccurate but for which it is not feasible to make full corrections at scale. With models of the second kind as introduced above, we can utilize limited samples of empirical data on error distributions to stochastically predict probable correct values for newly observed data. By repeated predictions of hypothetical correct values with random variation at the level of attributes of individual publications simulated distributions of indicator values can be calculated from the predicted basic data and summarized and expressed as uncertainty intervals for bibliometric indicator values, be it at the level of individual publications or arbitrary aggregates.

Above, we presented data on the uncertainty in citation count figures for one specific data source, WoS. There are other sources of uncertainty and in the following we will also consider one further source of uncertainty, namely errors in the document types of items assigned by bibliometric databases.

To study the effect of data inaccuracies, statistical models of the error distribution are required. Here we propose to use Bayesian regression, primarily because this allows the uncertainty in the fitted model parameters to affect the predicted values for new data. This is possible because, unlike in frequentist regression, the parameters in Bayesian regression are not modelled as scalar values but as probability distributions. We use the empirical data from the previous section\footnote{The citation count error models are estimated on empirical data of only a moderate sample size. The sample does not include sufficient data at large observed WoS citation counts, a consequence of the basic sampling strategy and the skewness of citation count data. We therefore restrict our simulations to moderate citation counts. The collected data can be extended with new sample data and new models estimated as the need for such modelling arises.}. This and all other models are independently estimated with the \texttt{R} package \texttt{brms}, v. 2.20, (\citeproc{ref-burkner2017brms}{Bürkner 2017}).

We estimated two independent Bayesian regression models. For the first one we estimated the number of missing citations of a publication from its observed citation count with a negative binomial model according to \(o_{i} \sim NegB(\ln (c_{i} + 1), \theta)\) and predicted error-free citation counts calculated as \(c^{* sim}_{i} = c^{new}_{i} + \hat{o_{i}}\). This notation follows Franceschini et al. (\citeproc{ref-franceschini2013novel}{2013}): in model estimation, \(o_{i}\) is the number of omitted citations and \(c_{i}\) the number of error-affected citations of a publication, collected from a sample of empirical data; \(\theta\) is the negative binomial overdispersion parameter, estimated from the data. In prediction from the estimated model, \(c^{* sim}_{i}\) is the simulated, i.e.~predicted, error-free number of citations which is calculated from observed new citation count data \(c^{new}_{i}\) and predicted missing citations \(\hat{o_{i}}\). A negative binomial model is used as it is suitable for non-negative and skewed count data. A log-transformation is applied to the independent variable as the otherwise excessive range would make the model unstable for high citation counts. To estimate the model, the data from section {[}\emph{Missing links: Empirical study of citation error distribution}{]} are used. Once this model is estimated, it can be used to stochastically predict the number of missing citations for any desired new input citation counts in a way that fully incorporates the available empirical knowledge, in this case the overall incidence of omitted citations and the relationship between omitted citations and total citation count.

In Bayesian modelling empirical knowledge beyond the data used to estimate the model can be included by specifying prior probability distributions. These reflect prior knowledge about model parameters and get updated via Bayes' rule with the data to obtain the posterior distribution. In the case of missing citations, some prior knowledge is available in the form of estimates of the rate of missing citations due to inexact reference matching reported in earlier studies. The results of Abt (\citeproc{ref-abt1992fraction}{1992}), Buchanan (\citeproc{ref-buchanan2006accuracy}{2006}), Franceschini et al. (\citeproc{ref-franceschini2013novel}{2013}), and Olensky et al. (\citeproc{ref-olensky2016evaluation}{2016}) suggest that the rate of missed citation in WoS can be expected to fall somewhere between 0 and 12 per cent. Assuming a plausible central value of 6 per cent and given that the observed average citation count in our sample is 16.4, this would indicate an average value of 1.0 missed citations with values above 0.0 and up to 2.0 still being plausible. From this we set a weakly informative prior distribution for the intercept term of Normal(0, 0.8) on the log scale. This sets the prior distribution on the original data scale centered on 1.0 citations with 95 per cent of the probability mass between 0.2 and 4.95 citations. As we have no particular previous knowledge on the influence of the predictor, we set a vague Normal(0, 1) prior for its coefficient. Prior predictive checks showed that these settings generate distributions broadly compatible with the data.

For the second model, the error-free document type is predicted with a multinomial regression model in which the error-affected observed document type is the single predictor variable according to \(d^{*}_{i} \sim categorical(d_{i})\). The empirical data to estimate this model is from Donner (\citeproc{ref-donner2017document}{2017}). In that study, the true document type was collected by inspection of the original document and assigned according to the published definitions of document types by the two databases which were analyzed. Both document type variables have four classes: article, review, letter, and other. Before we proceed with simulation exercises which investigate the influence of the two error sources we outline how inaccurate document type data can adversely affect bibliometric indicators.

First, bibliometric studies normally consider a subset of all publications of a unit of assessment. This subset is usually defined based on publication years and document types. Often the document types article, review, and letter are chosen and all others are not included in an analysis because not all types of documents are of interest in bibliometric research or evaluation. Hence, if the document type assignments are not correct, it is possible that not all relevant publications will be selected for analysis and that some irrelevant publications will be included, which can lead to incorrect publication counts. In this scenario, the citation counts of wrongly left out and wrongly included publications also distort citation impact indicator values.

The second and third kinds of error which can be caused by inaccurate document type data arise when constructing normalized citation impact indicators using reference sets to obtain reference values. The second error arises because the reference values for normalized citation indicators are also based on the selection of publications of specific types. For example, an observed citation count for a publication of type article is compared to the citation count of all publications of document type article in the same field and publication year. If, due to document type errors, the wrong publications are inadvertently used to calculate these citation score reference values, the calculation of normalized citation scores for the publications of interest in a study will also be affected.

The third kind of error is that when the document type assignment of an item is incorrect, the normalized citation scores of the target publications might get calculated on the basis of the wrong normalization value, namely that of the reference set for the incorrect document type.

\subsubsection*{Simulation exercise 1}\label{simulation-exercise-1}
\addcontentsline{toc}{subsubsection}{Simulation exercise 1}

We begin with a deliberately simple example of the proposed method. Subsequent simulations will be incrementally more complex in order to develop understanding of the approach step by step. In this basic example we look at the distributions of predicted error-free values of citation count and document type of three publications -- only these attributes at the micro-level of publications are examined and no indicators are calculated. We are given as observed, possibly error-affected data, publications P1 of type article with 5 citations, P2 of type review with 10 citations, and P3 of type letter with 0 citations. Using these observed values and the regression models described above, we predict the likely error-free values of their characteristics with 2000 draws from the posterior predictive distributions of the models for each publication and attribute. Table \ref{tab:m1-tab1} shows the predicted error-free document types of the three publications. According to these figures, it is unlikely but possible that P1 is a different type than article. In contrast, for P2 and P3 it is quite likely that the observed document types were inaccurate as substantial probabilities for types different than the observed ones are predicted. The figures for predicted error-free citations counts are shown in Table \ref{tab:m1-tab2}. In each of the three cases, the most likely citation count is the observed one but the numbers in the higher counts indicate that higher error-free citation counts are also consistent with the modelled empirical data and given observations. For example, for P2, the probability that the observed value of 10 citations is accurate is less than 70 per cent.

\begin{table}[!h]

\caption{\label{tab:m1-tab1}Simulation exercise 1. Predicted error-free document types of three publications}
\centering
\begin{tabular}[t]{lrrrr}
\toprule
publication & Article & Letter & Other & Review\\
\midrule
P1 & 1931 & 17 & 8 & 44\\
P2 & 253 & 0 & 55 & 1692\\
P3 & 726 & 1081 & 193 & 0\\
\bottomrule
\end{tabular}
\end{table}

\begin{table}[!h]

\caption{\label{tab:m1-tab2}Simulation exercise 1. Predicted error-free citation counts of three publications}
\centering
\resizebox{\linewidth}{!}{
\begin{tabular}[t]{lrrrrrrrrrrrrrrrrrrrrrrrr}
\toprule
publication & 0 & 1 & 2 & 3 & 4 & 5 & 6 & 7 & 8 & 9 & 10 & 11 & 12 & 13 & 14 & 15 & 16 & 17 & 18 & 19 & 20 & 21 & 22 & 23\\
\midrule
P1 & 0 & 0 & 0 & 0 & 0 & 1450 & 304 & 114 & 55 & 30 & 19 & 10 & 3 & 6 & 4 & 2 & 2 & 1 & 0 & 0 & 0 & 0 & 0 & 0\\
P2 & 0 & 0 & 0 & 0 & 0 & 0 & 0 & 0 & 0 & 0 & 1363 & 323 & 139 & 71 & 34 & 23 & 17 & 9 & 3 & 6 & 4 & 2 & 4 & 2\\
P3 & 1665 & 210 & 88 & 24 & 10 & 0 & 2 & 0 & 0 & 0 & 0 & 0 & 1 & 0 & 0 & 0 & 0 & 0 & 0 & 0 & 0 & 0 & 0 & 0\\
\bottomrule
\end{tabular}}
\end{table}

\subsubsection*{Simulation exercise 2}\label{simulation-exercise-2}
\addcontentsline{toc}{subsubsection}{Simulation exercise 2}

We now present a series of three simulations in which we first look at the effects of each of the two sources of data inaccuracies in isolation and then at their combined effect. To begin, in this exercise we look at citation count error only and assume that document types are free of error. To make the exercises both instructive and interesting, we model a basic evaluation scenario. Two research units with modest numbers of publications are evaluated with respect to their publication output and citation impact. The fundamental question in this scenario is whether, given the known data quality issues, the observed differences in scores are reliable reflections of actual performance differences or whether observed differences might be attributable to data inaccuracies alone.

For these next three exercises we first create a common data set of error-affected values for two research units, A and B, with 40 and 50 publications, respectively, of various document types. For calculating bibliometric indicators, only the subset of articles and reviews of those publications will be selected and counted (indicator publication count, P) while letters and items of other document types are discarded, as is common in bibliometric studies. Let the units only have published in one homogeneous research discipline and in the same publication year, so no further normalization for these factors needs to be applied. An additional reference set of publications consists of 200 publications of all document types. Document types of all three publication sets are sampled from the global document type distribution which, based on WoS data, is as follows: articles: 68 per cent, letters: 3 per cent, reviews: 4 per cent, other: 25 per cent. Citation counts of papers, the input quantity in this exercise, are generated from discretized lognormal distributions such that units A and B and the other publications each have their own typical citation impact levels, realized by setting the mean parameters of the distributions the citation count values are drawn from to different values (A: 0.8, B: 1.2, ref.: 1). Depending on the document type these parameters are multiplied with scaling factors as follows. Articles: 1; reviews: 1.5; letters: 0.2; other: 0.1.

For the two units A and B, the three basic bibliometric indicators, number of article and review publications (P), number of their citations (C), and mean normalized citation score (MNCS), are calculated. For the latter indicator, first, the item-level normalized citation scores (NCS) of the publications are calculated as the observed citation count divided by the expected citation count. The expected citation count is the average of the citation counts of all publications of the same document type. The MNCS of a unit is then computed as the arithmetic average of the NCS of its publications (\citeproc{ref-waltman2011towards}{Waltman et al. 2011}). The choice of indicators is purely illustrative and the present approach can be used for any other indicators.

For each of the observed citation count values of the whole publication set 2000 draws from the posterior predictive distribution of error-free citation count were drawn from the estimated citation count error Bayesian regression model. For each of these 2000 draws, the bibliometric indicator values for the two units were calculated. The distributions of these indicator values across the simulations are reported in summarized form in Table \ref{tab:sim2-results}. In this and the following similar tables the original fixed error-affected indicator scores are displayed in the left part and the values for the distributions of simulated error-free indicators in the right part of the tables. As document types were assumed to be error-free, the publication count indicator P is not subject to any variation. The units' total citation counts (C) and mean normalized citation score (MNCS) show clear differences in their median values as compared to the input values. Probable error-free citation counts are considerably greater than observed values would suggest. MNCS values for simulated error-free data are close to the observed values in the median of the distributions. For both indicators and both units the credible intervals conveying the information on the range of the values covering the central 95 per cent of the posterior probability distribution show that there is substantial uncertainty in these indicator results.

\begin{table}

\caption{\label{tab:sim2-results}Simulation exercise 2. Observed citation count error-affected and simulated error-free bibliometric indicator values}
\centering
\begin{tabular}[t]{lrrrlll}
\toprule
\multicolumn{1}{c}{ } & \multicolumn{3}{c}{observed error-affected values} & \multicolumn{3}{c}{simulated error-free values} \\
\cmidrule(l{3pt}r{3pt}){2-4} \cmidrule(l{3pt}r{3pt}){5-7}
publication set & P & C & MNCS & P & C & MNCS\\
\midrule
A & 31 & 76 & 0.55 & 31 (31, 31) & 88 (80, 103) & 0.58 (0.53, 0.67)\\
B & 38 & 185 & 1.07 & 38 (38, 38) & 204 (192, 222) & 1.06 (1, 1.14)\\
\bottomrule
\multicolumn{7}{l}{\rule{0pt}{1em}\textit{Note: }}\\
\multicolumn{7}{l}{\rule{0pt}{1em}Simulated values are distribution medians and 95\% credible intervals in parentheses}\\
\end{tabular}
\end{table}

\subsubsection*{Simulation exercise 3}\label{simulation-exercise-3}
\addcontentsline{toc}{subsubsection}{Simulation exercise 3}

In this exercise, the observed citation counts are assumed to be error-free while the document types are taken as being affected by data inaccuracies. Simulated error-free document types are drawn as predictions from the regression model estimated on the basis of actual document type error data from Donner (\citeproc{ref-donner2017document}{2017}). Again, 2000 draws from the Bayesian posterior predictive distribution are realized for each input datum. Different total citation counts and MNCS values can occur even though individual papers' citation counts are not altered in this scenario because different publications can be selected as relevant articles and reviews depending on the predicted document type. The results are shown Table \ref{tab:sim3-results}, where it can be seen that there is some uncertainty in publication counts, total citations, and consequently MNCS. For indicator C in this scenario simulated error-free values smaller than those for the observed data are also predicted. This is because in some cases highly cited documents of an included document type are predicted to be an excluded type and do not contribute to that score. Nonetheless, the median indicator values across all repetitions are close to their observed values and the uncertainties are of moderate magnitude.

\begin{table}

\caption{\label{tab:sim3-results}Simulation exercise 3. Observed document type error-affected and simulated error-free bibliometric indicator values}
\centering
\begin{tabular}[t]{lrrrlll}
\toprule
\multicolumn{1}{c}{ } & \multicolumn{3}{c}{observed error-affected values} & \multicolumn{3}{c}{simulated error-free values} \\
\cmidrule(l{3pt}r{3pt}){2-4} \cmidrule(l{3pt}r{3pt}){5-7}
publication set & P & C & MNCS & P & C & MNCS\\
\midrule
A & 31 & 76 & 0.55 & 32 (30, 34) & 78 (71, 85) & 0.57 (0.53, 0.61)\\
B & 38 & 185 & 1.07 & 39 (37, 42) & 185 (168, 189) & 1.07 (1.01, 1.14)\\
\bottomrule
\multicolumn{7}{l}{\rule{0pt}{1em}\textit{Note: }}\\
\multicolumn{7}{l}{\rule{0pt}{1em}Simulated values are distribution medians and 95\% credible intervals in parentheses}\\
\end{tabular}
\end{table}

\subsubsection*{Simulation exercise 4}\label{simulation-exercise-4}
\addcontentsline{toc}{subsubsection}{Simulation exercise 4}

\begin{table}

\caption{\label{tab:sim4-results}Simulation exercise 4. Observed error-affected and simulated error-free bibliometric indicator values}
\centering
\begin{tabular}[t]{lrrrlll}
\toprule
\multicolumn{1}{c}{ } & \multicolumn{3}{c}{observed error-affected values} & \multicolumn{3}{c}{simulated error-free values} \\
\cmidrule(l{3pt}r{3pt}){2-4} \cmidrule(l{3pt}r{3pt}){5-7}
publication set & P & C & MNCS & P & C & MNCS\\
\midrule
A & 31 & 76 & 0.55 & 32 (30, 34) & 91 (79, 108) & 0.59 (0.53, 0.68)\\
B & 38 & 185 & 1.07 & 39 (37, 41) & 204 (184, 223) & 1.07 (0.98, 1.17)\\
\bottomrule
\multicolumn{7}{l}{\rule{0pt}{1em}\textit{Note: }}\\
\multicolumn{7}{l}{\rule{0pt}{1em}Simulated values are distribution medians and 95\% credible intervals in parentheses}\\
\end{tabular}
\end{table}

As the final exercise in this series with the same initial observed data we specify that both document type and citation count are affected by realistic levels of inaccuracy. To this end, predicted values for both kinds of data are drawn from the respective regression models as specified above. Results for 2000 simulation runs are displayed in Table \ref{tab:sim4-results}. The simulation shows that in direct consequence of the empirical data and model structure the observed citation sums are on average underestimated, but also citation sums smaller than the observed ones are consistent with the error distributions. This is due the uncertainty in document types and the inclusion of specific document types. While in the observed data there are 38 articles and reviews for unit B, the simulations show that there would likely be between 37 and 41 such publications with 95 per cent posterior probability if we had error-free data. The results further indicate that the mean normalized citation score, absent errors, would be between 0.53 and 0.68 for unit A and between 0.98 and 1.17 for unit B, with 95 per cent probability each.

\subsection{A real-world example application}\label{a-real-world-example-application}

Up to this point, only synthetic exemplary data was utilized. The question remains, does any of this matter in real-world scenarios? To find out, in this section real bibliometric data on the level of formal research groups is analysed.

Data for all WoS-indexed publications of 110 chemistry research groups of six universities in one German federal state was collected in the course of a commissioned provision of bibliometric data in support of a state-wide research evaluation in the discipline of chemistry by an expert panel. Universities and research groups are kept unidentified in this analysis. The data set comprises 3818 different publication records. The groups were asked to submit article and review publications from 2008 to 2013 and for those years there are between about 400 and 700 publications per year, and a smaller number of publications in 2007 and in 2014. The groups published between 3 and 280 papers. Several groups have small publication counts as they were established during the observation period and have only been active for a short period. In order to calculate field-normalized citation impact scores, we use CWTS's algorithmically constructed classification system, as first described in Waltman \& Van Eck (\citeproc{ref-waltman2012new}{2012}). Specifically, we use the micro-level fields of the 2020 version of the system, which are the outputs of clustering the citation network with the `Leiden algorithm' (\citeproc{ref-traag2019louvain}{Traag et al. 2019}). The publications are distributed over 556 out of 4013 micro fields of that system\footnote{It should be mentioned that the choice of classification system, insofar as it is used for calculating normalized citation scores, can be seen as a source of uncertainty for such indicators, albeit one not based on data inaccuracies. Different classification systems have been shown to produce different item-level normalized citation scores (\citeproc{ref-haunschild2022scores}{Haunschild et al. 2022}; \citeproc{ref-robinson2023errors}{Robinson-Garcia et al. 2023}; \citeproc{ref-ruiz2015field-normalized}{Ruiz-Castillo \& Waltman 2015}; \citeproc{ref-waltman2019field}{Waltman \& Van Eck 2019}). It would be difficult to argue that any one particular classification system of the many possible alternatives that could reasonably be used is the most suitable one. Furthermore, some algorithmic classification methods have a random component and repeated runs on the same data with the same parameters can lead to more or less different clustering solutions, the Leiden algorithm being one example. Nonetheless, the choice of this classification system for the present study is deliberate. Unlike the vendor-supplied classification system of WoS, the Leiden system is applied at the level of individual publications, rather than entire journals. Papers of one journal can therefore be classified into different micro-level fields (ideally ``topics''), which is highly appropriate for all but the most specialized small journals. Moreover, the Leiden system has the practical advantage that each paper is assigned to exactly one micro field. This should not be underestimated. It eliminates calculating scores of publications for each classification class and then aggregating them, which is required when papers are assigned multiple classes. While algorithmically constructed classification systems are currently not able to capture human-conceived topics perfectly (\citeproc{ref-haunschild2018algorithmically}{Haunschild et al. 2018}; \citeproc{ref-held2021challenges}{Held et al. 2021}; \citeproc{ref-waltman2012new}{Waltman \& Van Eck 2012}), recent empirical research has confirmed that publication-level classification systems, including algorithmic ones, are more accurate than journal-level systems (\citeproc{ref-colliander2019comparison}{Colliander \& Ahlgren 2019}; \citeproc{ref-klavans2017type}{Klavans \& Boyack 2017}; \citeproc{ref-ruiz2015field-normalized}{Ruiz-Castillo \& Waltman 2015}; \citeproc{ref-shu2019comparing}{Shu et al. 2019}).}. There were no micro level field assignments for 79 records and these are therefore excluded from further analysis\footnote{These excluded records were, by WoS document type: 43 editorial material, 22 meeting abstracts, 4 news items, 3 biographical items, 3 corrections, 2 articles, 1 letter, and 1 reprint. The two article type records contained no references.}. For each combination of publication year and micro class which contains a publication from the chemistry groups we also retrieved all other WoS publication records in order to construct complete reference sets, including citation counts up to spring 2020. This larger set includes 863,443 publication records. Next, document types were simplified by designating all types other than article, review, and letter as `other', so that there are four categories.

As all data are obtained from WoS, we assume them to be subject to the empirically found levels of data inaccuracies and we use the statistical models estimated from original WoS data for posterior predictive inference to obtain likely correct versions of the data on document type and complete citation count, i.e.~the models of the second kind from the preceding section. The complete citation count is the WoS citation count plus the predicted number of missed citations from the regression model. The simulated correct document types are also predicted based on the observed WoS document types using the estimated regression model. Because of the larger size of the data set we were limited to creating 1000 estimated data sets of predicted error-free values. This figure is large enough to obtain stable and reliable results.

\begin{figure}
\centering
\includegraphics{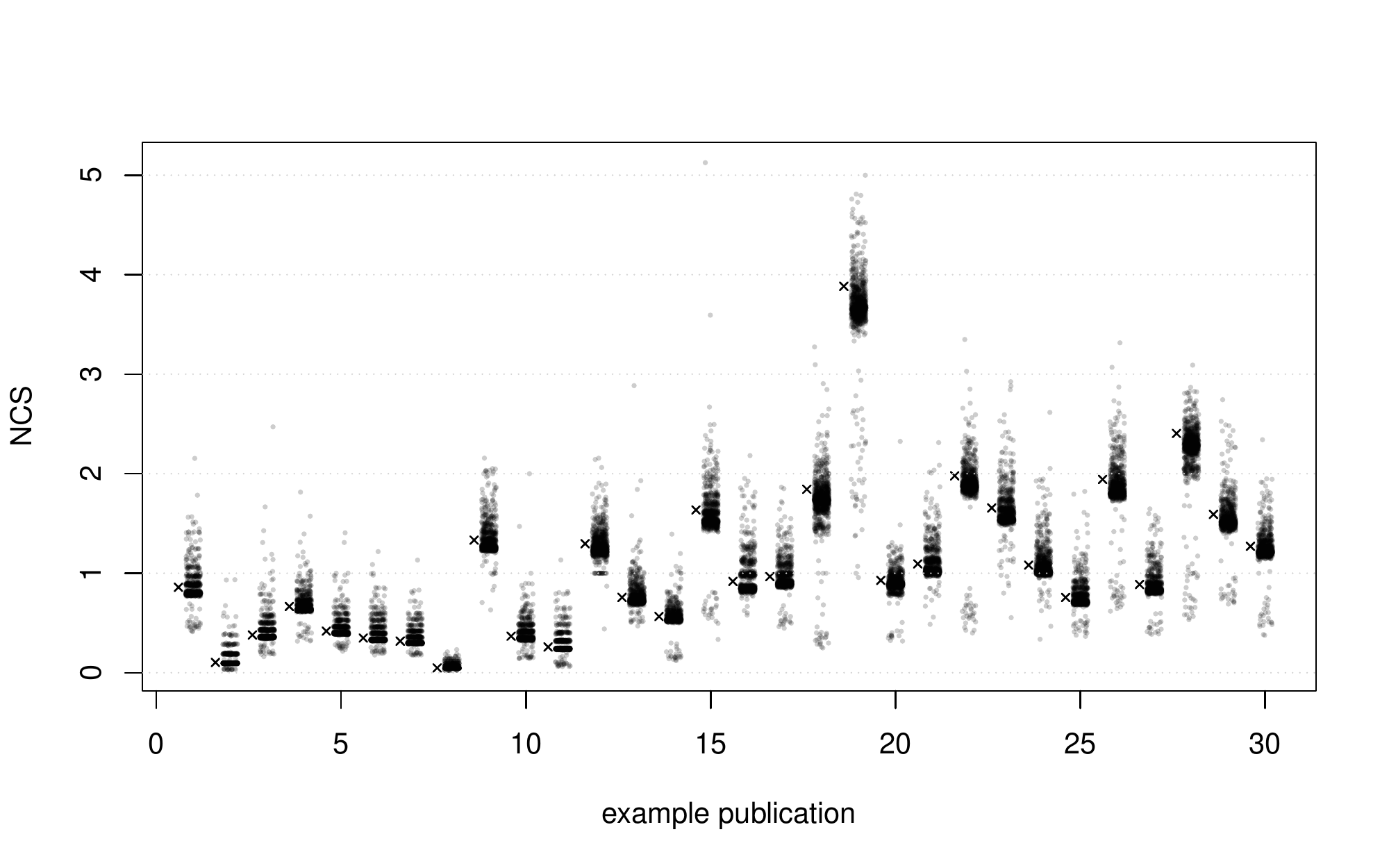}
\caption{\label{fig:item-uncertainty-plot}Estimates of normalized citations scores for a sample of publications, 1000 runs, symbols: \(\times\) original error-affected data, \(\bullet\) simulated error-free data}
\end{figure}

Figure \ref{fig:item-uncertainty-plot} shows the distributions of normalized citation scores (observed citations divided by average citations of publication of the same document type category, publication year, and micro-level field) for a sample of publications of the chemistry groups data set. Grey dots are simulated error-corrected values and \(\times\) symbols are the uncorrected observed values. While all example items have a central zone in which most of the predictions are concentrated, they also all show substantial item-level variability across Monte Carlo runs. This illustrates the uncertainty propagated from data errors in citation matching and document type assignment to item-level bibliometric indicator scores.

\begin{figure}
\centering
\includegraphics{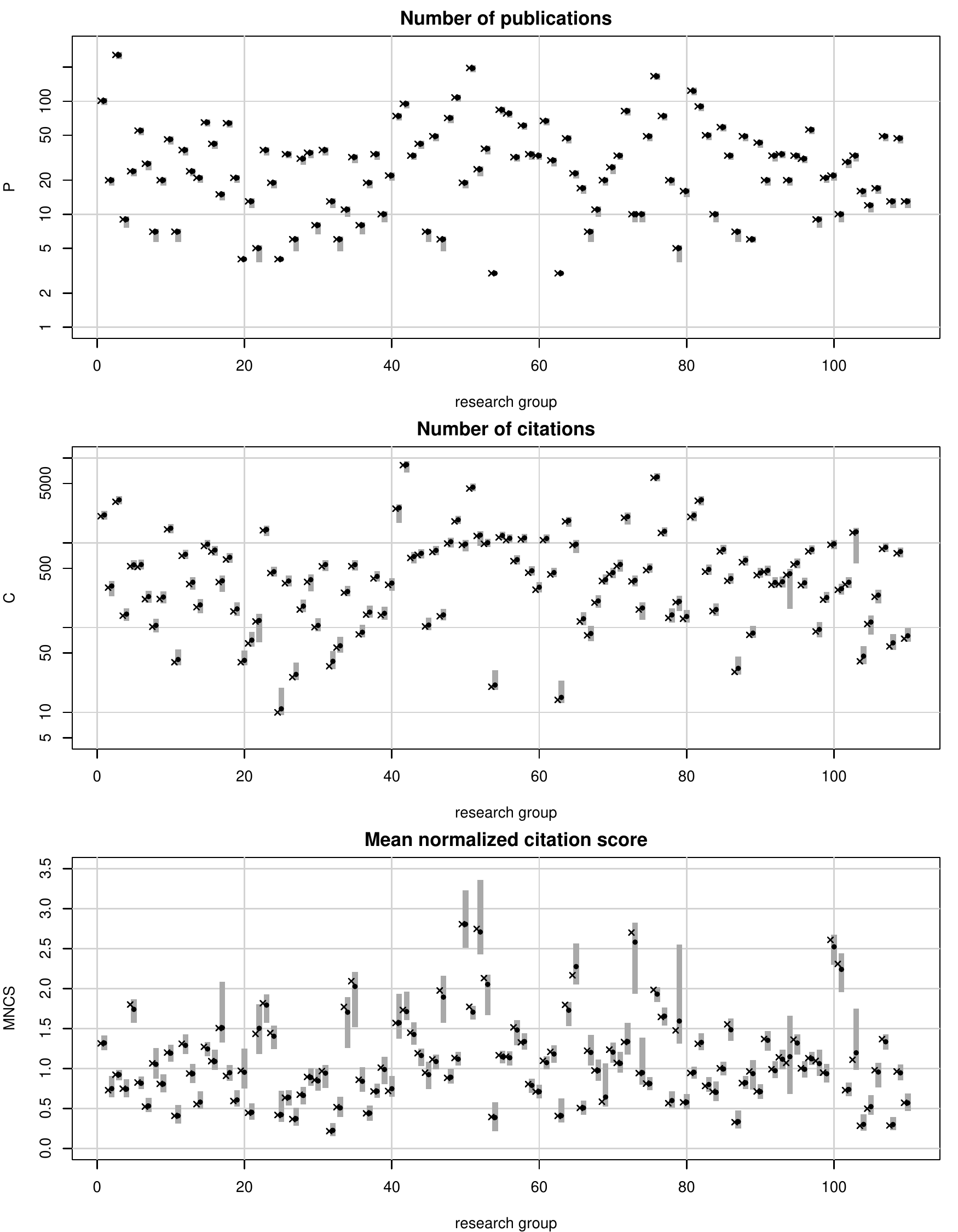}
\caption{\label{fig:chem-groups-plot}Estimated error-corrected data for publications (P), citations (C), and mean normalized citation score (MNCS) for 110 chemistry research groups, medians and 95\% credible intervals, 1000 runs, symbols: \(\times\) original error-affected data, \(\bullet\) simulated error-free data}
\end{figure}

Figure \ref{fig:chem-groups-plot} shows the uncertainty in bibliometric indicator values on the aggregated level of research groups. The plots show black circles for the medians of the posterior predictive distribution of the indicator values and 95 per cent credible intervals plotted as gray bars for the uncertainty in these estimates, which cover the central 95 per cent of the simulated values. The \(\times\) symbols show the uncorrected values. The number of publications typically exhibits relative uncertainty of around 6 per cent on average, computed as the percent ratio of the difference of the bounds of the credible interval to the value of the median, the best estimate for the error-free value. The values of research groups with small numbers of publications are barely affected by uncertainty, as they have very low absolute chance of being affected by document type assignment errors. The largest calculated credible interval is for the group with the greatest number of observed publications. This group reported 257 publications and its 95 per cent credible interval is 250 to 257 with a best estimate (median) of 256.

Relative uncertainty in group citation sums is greater, being around 18 per cent on average. Readers should bear in mind that the y-axes on these two plots are logarithmically scaled to better show the whole range of values, which contracts the apparent size of uncertainty bars in the higher values. Uncertainties are quite large for groups in the whole range of total citations. The group with the smallest reported value of 10 citations has a credible interval from 10 to 17 citations, or 70 per cent. On the opposite end of the spectrum, the one with the largest reported value of 8239 citations has a credible interval from 7219 to 8490, or 15 per cent relative uncertainty.

The relative uncertainty in the group level estimates of MNCS is on average 23 per cent. Figure \ref{fig:chem-groups-plot}, bottom panel, clearly shows that higher values are associated with greater uncertainties. The largest uncertainty is found for a group with a best estimate MNCS of 1.6, with credible interval of 1.3 to 2.5, while its uncorrected MNCS value is 1.5 -- a 79 per cent relative uncertainty. Best estimates of error-free MNCS values are typically close to the observed error-affected values but associated with large uncertainty intervals.

\begin{figure}
\centering
\includegraphics{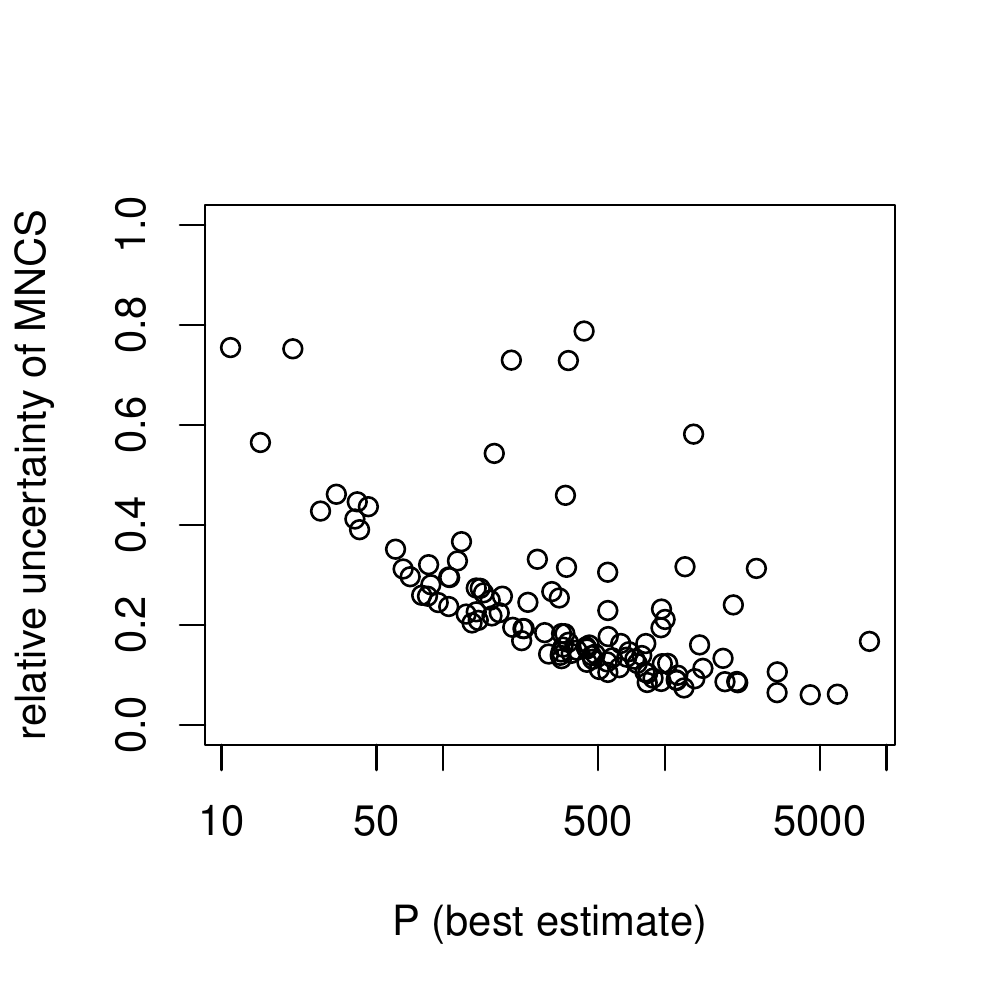}
\caption{\label{fig:chem-groups-size-unc-plot}Relationship between P and uncertainty in MNCS, 110 chemistry research groups}
\end{figure}

Figure \ref{fig:chem-groups-size-unc-plot} shows the relationship between the best estimate of the publication count, P, and the relative uncertainty in the MNCS. Each data point is one research group. There is a clear and robust association, groups with fewer publications have MNCS values associated with greater uncertainties. However, this relationship is far from perfectly log-linear, as there are a number of groups with much higher uncertainties than others with similar publication counts. Even for groups with hundreds of publications, the relative uncertainty can be on the order of 80 per cent.

\section{Discussion}\label{discussion}

\subsection{Summary}\label{summary}

In this study we have outlined a general approach of the estimation of uncertainty in bibliometric indicators. It consists of using empirical data on item-level error distributions for bibliometric quantities to estimate Bayesian regression models and generating predictions from the model for new data as Monte Carlo simulations. This allows the propagation of input uncertainties whose distributions are known to any bibliometric indicators of interest. Thus the uncertainty in bibliometric indicators that is due to data quality issues can be explicitly modelled and expressed. This uncertainty can be expressed, for example, as median estimates of a quantity and corresponding credible intervals corresponding to a chosen probability that the likely correct value lies within that range. We have demonstrated this approach with a number of simple synthetic examples for didactic purposes and for one example of real-world data with a specific evaluative bibliometric application. These examples showed substantial uncertainty in bibliometric indicator values that would otherwise have have remained invisible.

\subsection{Limitations and future work}\label{limitations-and-future-work}

The present study has several limitations. We have worked only with one moderately sized manually assessed data set for the validation of citation counts from a single data source. This is clearly not generalizable to every scenario but can at best serve as one example to be extended and supplemented with further samples in the future. Our results do not transfer to citation counts from any other data source or even the same data source once the proprietary algorithm for reference matching is changed substantially. As a case in point, the study by Van Eck \& Waltman (\citeproc{ref-vaneck2017accuracy}{2017}) shows that the types of errors in citation data differ by data source. Furthermore, we have used a single predictor variable, WoS observed citation counts, to predict omitted citations. This implies the assumption that the number of omitted citations is otherwise not affected by any characteristics of the specific publication. This is a necessary simplification due to constraints in data collection. There are other relevant predictor variables for omitted citations. One example would be the scientific field of a publication. For instance, in the humanities, complex footnote citations are common, as opposed to reference lists. Such citation data is inherently more difficult to extract correctly and completely, which could manifest in greater citation error in this domain, as results in Tüür-Fröhlich (\citeproc{ref-tuur2016non}{2016}) suggest. Another example for a possibly useful predictor variable would be publication year. Publishing practices and data collection practices of citation index databases change over time. Therefore data originating in different time periods might be differently affected by citation count errors. A third example would be publications from languages other than English being cited as references. These may possibly exhibit greater error rates than English-language references if data extraction methods are primarily designed for English language data and perform worse for other languages. In many cases such cited references are also cited in translated or transcribed form, which are practices susceptible to the introduction of errors or variability, such as when competing romanization systems are used for non-Latin scripts. Moreover, the research of Moed \& Vriens (\citeproc{ref-moed1989possible}{1989}) and Buchanan (\citeproc{ref-buchanan2006accuracy}{2006}) showed that specific journals are more affected by referencing errors, as are papers by authors with complex names. Including such other predictors as covariates into regression models like the ones discussed here would require larger samples sizes in order to achieve reliable estimates across ranges of numerical and categories of categorical variables.

Another limitation is the choice of specification for the regression models used in the stochastic prediction of correct and error-affected data. For the citation error model, we chose a negative binomial model as it is a count data model well suited to skew data such as citation counts. Other choices are possible and perhaps better. The purpose of the present study was not to identify the single best specification choice, for that the size of the sample was not large enough, but rather to outline one approach and test it for one reasonable model choice. Moreover, incorrect document types and incomplete citation counts are just two of a larger number of known data error-related phenomena. We have focused on omitted citations, a source of uncertainty which is documented and understood relatively well and is assumed to have considerable effects on data quality. Thus we have not taken into account other phenomena of citation error such as phantom citations or unindexed citations from supplemental material. Ultimately all sources of data error need to be either eliminated by correction or included in a more comprehensive error assessment of bibliometric indicator results. We expect that increasing the sample sizes, including additional predictors and using an alternative regression model specification would lead only to minor quantitative changes in the results we report here. On the other hand, the inclusion of more types of data errors can be expected to lead to increases in predicted uncertainties. In this sense, the present limitations mean that our reported uncertainty intervals are probably conservatively small.

We have restricted the study to three very basic bibliometric indicators, publication counts, citation counts, and mean normalized citation scores. While most bibliometric studies use the first two indicators, the MNCS indicator is just one of the many possible choices to characterize the typical citation impact of a set of publications. It is known that this indicator can be sensitive to influence by a few very highly cited papers. A more comprehensive study of the influence of data errors on other important impact indicators, including representatives of other classes of citation impact indicators, such as highly cited rate indicators and indicators of total impact, is certainly warranted. It might be prudent to investigate if any citation impact indicators are particularly volatile or robust to data errors.

This study only looked at one kind of research unit when studying real-world data, formal university research groups. Again, this example was chosen for being illustrative. It is arguably the smallest type of unit that is reasonably amenable to bibliometric scrutiny, as individual researchers do not commonly have the necessary number of publications to make statistical analysis reliable. This is not even strictly the case for the research groups used here. The intention behind the choice of this small a unit was to make any possibly found uncertainties easy to perceive. It remains to be studied if the quite substantial uncertainties found here can also be seen in larger units, such as departments or whole organizations, or if they might become small for large enough publication sets. In any case, the uncertainties at the level of single publications will not disappear until the data are free of any error.

\subsection{Conclusion}\label{conclusion}

In the first empirical part of this work we studied the incidence of missing citation links in WoS. According to a global random sample, about 4 per cent of citation links are missing in WoS and nearly one third of all studied items had one or more missing citations. The citation error distribution was found to be highly skewed. It should be pointed out that citation index providers continue to improve their databases including their original data indexing quality and their reference matching procedures, so our results represent a snapshot of data quality at one particular point in time. Furthermore, both WoS and Scopus welcome corrections and incorporate them into their data after confirmation. So any discovered errors or omissions could in principle be corrected in the data source.

In the second empirical part we used these data and data from an earlier study to statistically model the errors in citation counts and document type assignments. We demonstrated how such models can be used to simulate error-corrected data from typical error-affected real data to gain insight into the range of bibliometric indicator values consistent with the data at hand and known error distributions. This method has the potential to improve the reporting of bibliometric indicator values and thus the interpretation of the results of bibliometric studies. It provides one practical implementation of the eighth principle of the Leiden Manifesto, which states that information on uncertainty and error in values should be quantified and reported alongside indicator values whenever possible.

By studying real bibliometric data that was used in assessment of university research groups in chemistry, we were able to show that data errors do not just cancel out and vanish at this level of aggregation but lead to quite substantial uncertainties in the values of bibliometric indicators. The hitherto indifferent attitude of professional bibliometricians to data quality and data errors can hardly be sustained in the light of the results presented herein. Uncertainty quantification can contribute to the legitimacy of the field of scientometrics because it increases transparency about known inaccuracies in data, instead of ignoring them.

\textbf{Acknowledgements}\\
I want to thank Nees Jan van Eck for in-depth discussion of the manuscript and for providing the algorithmic classification data. I also thank Stephan Stahlschmidt for helpful feedback on the manuscript, Anastasiia Tcypina for help in collecting the empirical data, and Sybille Hinze and Stefan Hornbostel for first introducing me to the concept of `calculus of errors' for bibliometrics. I am also grateful for the thoughtful suggestions of two anonymous referees.

\section{Appendix}\label{appendix}

\subsection{Simulating error-affected data}\label{simulating-error-affected-data}

In this section, we demonstrate the use of simulation of realistic errors for given error-free data sets. These are the models of the first kind of the section {[}\emph{Incorporating information about uncertainty into statistical models}{]}. They are probably of more limited practical use but very instructive nevertheless. For reasons of exposition, we will begin with a small example data set and consider the two sources of data error, omitted citations and inaccurate document types, first in isolation and then in combination.

The number of missed citations, a count variable, is modelled using a negative binomial regression model with a single predictor, the accurate number of citations. This predictor is modified by adding 1 and taking the natural logarithm: \(o_{i} \sim NegB(\ln( c^*_{i}+1), \theta)\). This notation also follows Franceschini et al. (\citeproc{ref-franceschini2013novel}{2013}): \(o_{i}\) is the number of omitted citations, \(c^*_{i}\) is the accurate number of citations; \(\theta\) is the negative binomial overdispersion parameter, estimated from the data.

The simulated error-affected citation count can then be obtained by subtracting the predictions from the model for new data from the input values. A drawback of the particular model is that it can predict numbers of missed citations greater than the number of true citations. This would result in predicted negative citation counts which is incongruous with reality. We therefore set a lower limit of 0 on the simulated citation counts: \(c^{sim}_{i}=c^{new}_{i} - \hat{o_{i}} \; \; \;\text{ subject to: } c^{sim}_{i}\ge 0\).

\subsubsection*{Simulation exercise A1}\label{simulation-exercise-a1}
\addcontentsline{toc}{subsubsection}{Simulation exercise A1}

We generate a synthetic data set of known accurate values for two research units, A and B, with 20 and 30 publications, respectively, of various document types. For calculating bibliometric indicators, only articles and reviews will be selected and counted (indicator publication count, P) while letters and items of other document types are discarded, as is common in bibliometric studies. The units have only published in one homogeneous research discipline and in the same publication year, so no further normalization for such factors needs to be applied. An additional reference set of publications consists of 1000 publications of all document types. Document types of all three publication sets are sampled from the global document type distribution which, based on WoS data, is as follows: articles: 68 per cent, letters: 3 per cent, reviews: 4 per cent, other: 25 per cent. Citation counts of papers, the input quantity in this exercise, are generated from discretized lognormal distributions such that units A and B and the other publications each have their own typical citation impact levels, realized by setting the mean parameters of the distributions the citation count values are drawn from at different values (A: ln 5, B: ln 7, ref.: ln 6). Depending on the document type these parameters are scaled with multiplication factors as follows, with articles at 1.0: reviews: 1.5, letters: 0.2, other: 0.1. For the two units, the three basic bibliometric indicators number of article and review publications (P), number of citations of these (C), and mean normalized citation score (MNCS), are calculated. The item-level normalized citation score (NCS) of a publication is calculated as the observed citation count divided by the expected citation count. The expected citation count is the average of the citation counts of all publications of the same document type. The MNCS of a unit is the arithmetic average of the NCS of its publications (\citeproc{ref-waltman2011towards}{Waltman et al. 2011}). The choice of indicators is purely illustrative and the present approach can be used for any other indicators.

The generated synthetic correct values are used as inputs to the statistical error model of the first kind for citation counts, which was estimated using the empirical data from the first part of the paper. In this exercise, no error is assumed for document types, so that data is unchanged. From the model, 2000 samples of the posterior predictive distribution of the omitted citation count for each publication item are drawn. In each case, the simulated error-affected citation count of a paper is calculated by subtracting the predicted number of omitted citations from the error-free original citation count. For each such set of simulated error-affected data the bibliometric indicators (P, C, MNCS) of the two units are calculated. By this method we can obtain distributions of the indicator quantities of interest generated from Bayesian posterior predictive distribution samples, given the known error-free data and the statistical model estimated using empirical data. In other words, distributions of the probable data that might have been available from WoS given the knowledge of the error distributions in this data source and the values of the input quantities.

\begin{figure}
\centering
\includegraphics{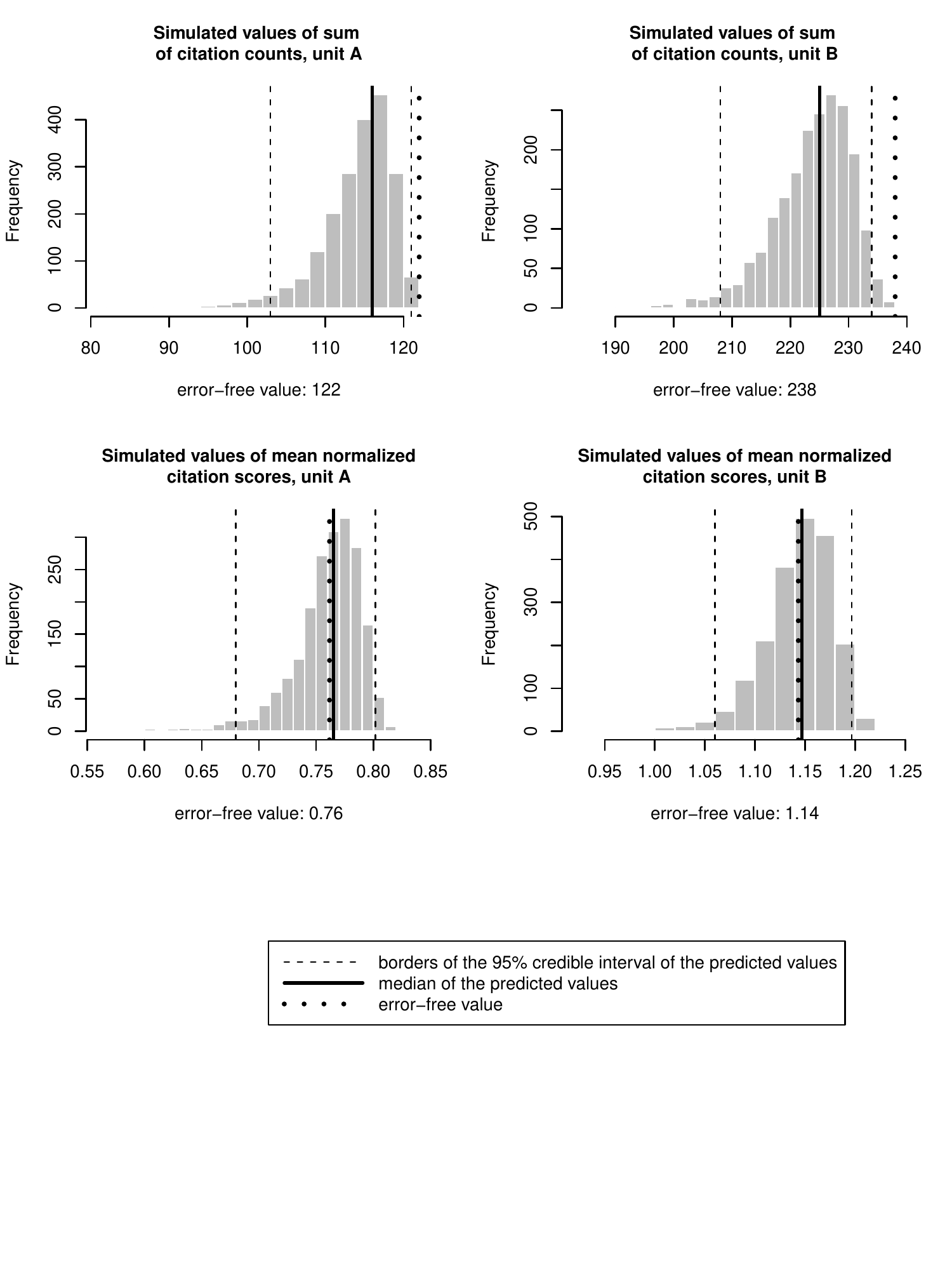}
\caption{\label{fig:sim-01-res}Results simulation exercise A1 - data affected by citation count error}
\end{figure}

The correct values and summaries of the obtained distributions of simulated error-affected values for the bibliometric indicators are reported in Table \ref{tab:sim-01-values}. There is no variability due to document assignment error, so the document types and thus the values for P are constant. Citation counts in the simulated inaccurate data are lower than the correct figures. Median simulated MNCS estimates are not far off their error-free values but the uncertainty is substantial. Appendix Figure \ref{fig:sim-01-res} shows the results for 2000 repetitions.

\begin{table}

\caption{\label{tab:sim-01-values}Simulation exercise A1. Error-free and simulated citation count error-affected bibliometric indicator values}
\centering
\begin{tabular}[t]{lrrrlll}
\toprule
\multicolumn{1}{c}{ } & \multicolumn{3}{c}{error-free values} & \multicolumn{3}{c}{simulated citation count error-affected values} \\
\cmidrule(l{3pt}r{3pt}){2-4} \cmidrule(l{3pt}r{3pt}){5-7}
publication set & P & C & MNCS & P & C & MNCS\\
\midrule
A & 16 & 122 & 0.76 & 16 (16, 16) & 116 (102.975, 121) & 0.77 (0.68, 0.8)\\
B & 20 & 238 & 1.14 & 20 (20, 20) & 225 (208, 234) & 1.15 (1.06, 1.2)\\
\bottomrule
\multicolumn{7}{l}{\rule{0pt}{1em}\textit{Note: }}\\
\multicolumn{7}{l}{\rule{0pt}{1em}Simulated values are distribution medians and 95\% credible intervals in parentheses}\\
\end{tabular}
\end{table}

\subsubsection*{Simulation exercise A2}\label{simulation-exercise-a2}
\addcontentsline{toc}{subsubsection}{Simulation exercise A2}

In this exercise, we look at the isolated effect of error-affected document type data on bibliometric indicator uncertainty. Citation counts are not changed from their original error-free values.

For the model of error-affected document types, we create a categorical distribution regression model that predicts the possibly error-affected document type \(d_{i}\) (four classes) assigned by the database vendor from a single predictor, the known correct document type \(d_{i}^{*}\) (same four classes): \(d_{i} \sim categorical(d_{i}^{*})\). The model is estimated with the empirical data on document type inaccuracy in WoS from Donner (\citeproc{ref-donner2017document}{2017}). The estimated model is used to run a simulation to illustrate the influence of realistic document type error using the setting of the prior exercise. In contrast to that exercise, here the citation counts of items are not affected by any error and thus constant. Again, 2000 sets of values are simulated and the three basic indicators calculated. The results are shown in Appendix Table \ref{tab:sim-02-values}. It can be seen that all aggregate level indicators are affected by document type errors: publication count because different numbers of documents with the document types of interest are counted and citation count for the same reason. MNCS is affected in addition to that because of distorted reference group citation count values of document types. We can see that indicator value uncertainties in this exercise are greater than in the previous one in which only citation counts were affected by errors.

\begin{table}

\caption{\label{tab:sim-02-values}Simulation exercise A2. Error-free and simulated document type error-affected bibliometric indicator values}
\centering
\begin{tabular}[t]{lrrrlll}
\toprule
\multicolumn{1}{c}{ } & \multicolumn{3}{c}{error-free values} & \multicolumn{3}{c}{simulated document type error-affected values} \\
\cmidrule(l{3pt}r{3pt}){2-4} \cmidrule(l{3pt}r{3pt}){5-7}
publication set & P & C & MNCS & P & C & MNCS\\
\midrule
A & 16 & 122 & 0.76 & 12 (8, 15) & 105 (42, 119) & 0.84 (0.37, 1.13)\\
B & 20 & 238 & 1.14 & 15 (11, 19) & 185 (122, 231) & 1.16 (0.83, 1.45)\\
\bottomrule
\multicolumn{7}{l}{\rule{0pt}{1em}\textit{Note: }}\\
\multicolumn{7}{l}{\rule{0pt}{1em}Simulated values are distribution medians and 95\% credible intervals in parentheses}\\
\end{tabular}
\end{table}

\subsubsection*{Simulation exercise A3}\label{simulation-exercise-a3}
\addcontentsline{toc}{subsubsection}{Simulation exercise A3}

What happens if both types of errors are present? This can be studied by simulating error-affected values from both the document type error model and the citation count error model. Using the same input data as before, the results are shown in Appendix Table \ref{tab:sim-03-values}. The results are comparable in uncertainty to those from the simulation exercise in which only document type error was modelled. In this particular scenario the contribution of document type error to overall uncertainty in bibliometric indicators is clearly greater than that of citation count error. Such a finding could only be uncovered by this modelling-prediction-simulation approach.

\begin{table}

\caption{\label{tab:sim-03-values}Simulation exercise A3. Error-free and simulated citation count error and document type error-affected bibliometric indicator values}
\centering
\begin{tabular}[t]{lrrrlll}
\toprule
\multicolumn{1}{c}{ } & \multicolumn{3}{c}{error-free values} & \multicolumn{3}{c}{simulated citation count error- and document type error-affected values} \\
\cmidrule(l{3pt}r{3pt}){2-4} \cmidrule(l{3pt}r{3pt}){5-7}
publication set & P & C & MNCS & P & C & MNCS\\
\midrule
A & 16 & 122 & 0.76 & 12 (8, 15) & 98 (38, 115) & 0.84 (0.36, 1.15)\\
B & 20 & 238 & 1.14 & 15 (12, 19) & 176 (115, 222) & 1.15 (0.83, 1.45)\\
\bottomrule
\multicolumn{7}{l}{\rule{0pt}{1em}\textit{Note: }}\\
\multicolumn{7}{l}{\rule{0pt}{1em}Simulated values are distribution medians and 95\% credible intervals in parentheses}\\
\end{tabular}
\end{table}

\subsubsection*{Simulation exercise A4}\label{simulation-exercise-a4}
\addcontentsline{toc}{subsubsection}{Simulation exercise A4}

The first set of three scenarios had units with deliberately small publication counts to support conveying the basic approach. At such low sample sizes the effects of errors are easy to observe from the large variability they introduce. However, this variability could conceivably be a minor factor in the much larger sample sizes which are common in bibliometrics. To test this hypothesis, we now investigate what happens if the publication counts are much higher. We keep all settings as before but increase the number of simulated publications for all three publication sets, A, B, and ref., by a factor of 100 to 2000, 3000, and 10,000 publications, respectively. The results are reported in Appendix Table \ref{tab:simulationA4}. The accuracy of the simulated error-affected MNCS indicator value has now improved a lot, yet there is still an appreciable uncertainty. Moreover, the possible error-affected values for P and C exhibit a wide range of results in absolute terms and are not close to their error-free values.

\begin{table}

\caption{\label{tab:simulationA4}Simulation exercise A4. Error-free and simulated error-affected bibliometric indicator values}
\centering
\begin{tabular}[t]{lrrrlll}
\toprule
\multicolumn{1}{c}{ } & \multicolumn{3}{c}{error-free values} & \multicolumn{3}{c}{simulated error-affected values} \\
\cmidrule(l{3pt}r{3pt}){2-4} \cmidrule(l{3pt}r{3pt}){5-7}
publication set & P & C & MNCS & P & C & MNCS\\
\midrule
A & 1472 & 12813 & 0.84 & 1090 (1005, 1169) & 9009 (8273, 9725) & 0.84 (0.8, 0.87)\\
B & 2202 & 26977 & 1.19 & 1628 (1504, 1734) & 19057 (17553, 20268) & 1.19 (1.15, 1.22)\\
\bottomrule
\multicolumn{7}{l}{\rule{0pt}{1em}\textit{Note: }}\\
\multicolumn{7}{l}{\rule{0pt}{1em}Simulated values are distribution medians and 95 per cent credible intervals in parentheses}\\
\end{tabular}
\end{table}

These results show that simulating realistic errors with different conditions can be informative about the contribution of different types of errors to overall indicator value uncertainties and about how the accuracy of indicator results changes with sample size.

\section*{References}\label{references}
\addcontentsline{toc}{section}{References}

\phantomsection\label{refs}


\begin{CSLReferences}{1}{0}
\bibitem[\citeproctext]{ref-abt1992fraction}
Abt, H. A. (1992). What fraction of literature references are incorrect? \emph{Publications of the Astronomical Society of the Pacific}, 104/673: 235. DOI: \href{https://doi.org/10.1086/132982}{10.1086/132982}

\bibitem[\citeproctext]{ref-alvarez2017funding}
Álvarez-Bornstein, B., Morillo, F., \& Bordons, M. (2017). Funding acknowledgments in the {Web of Science}: Completeness and accuracy of collected data. \emph{Scientometrics}, 112/3: 1793--812. DOI: \href{https://doi.org/10.1007/s11192-017-2453-4}{10.1007/s11192-017-2453-4}

\bibitem[\citeproctext]{ref-baas2020scopus}
Baas, J., Schotten, M., Plume, A., Côté, G., \& Karimi, R. (2020). Scopus as a curated, high-quality bibliometric data source for academic research in quantitative science studies. \emph{Quantitative science studies}, 1/1: 377--86. DOI: \href{https://doi.org/10.1162/qss_a_00019}{10.1162/qss\_a\_00019}

\bibitem[\citeproctext]{ref-birkle2020web}
Birkle, C., Pendlebury, D. A., Schnell, J., \& Adams, J. (2020). {Web of Science as a data source for research on scientific and scholarly activity}. \emph{Quantitative Science Studies}, 1/1: 363--76. DOI: \href{https://doi.org/10.1162/qss_a_00018}{10.1162/qss\_a\_00018}

\bibitem[\citeproctext]{ref-buchanan2006accuracy}
Buchanan, R. A. (2006). Accuracy of cited references: The role of citation databases. \emph{College \& Research Libraries}, 67/4: 292--303. DOI: \href{https://doi.org/10.5860/crl.67.4.292}{10.5860/crl.67.4.292}

\bibitem[\citeproctext]{ref-burkner2017brms}
Bürkner, P.-C. (2017). {brms}: An {R} package for {Bayesian} generalized linear mixed models using {Stan}. \emph{Journal of Statistical Software}, 80/1. DOI: \href{https://doi.org/10.18637/jss.v080.i01}{10.18637/jss.v080.i01}

\bibitem[\citeproctext]{ref-cioffi2022identifying}
Cioffi, A., Coppini, S., Massari, A., Moretti, A., Peroni, S., Santini, C., \& Shahidzadeh Asadi, N. (2022). {Identifying and correcting invalid citations due to DOI errors in Crossref data}. \emph{Scientometrics}, 127/6: 3593--612. DOI: \href{https://doi.org/10.1007/s11192-022-04367-w}{10.1007/s11192-022-04367-w}

\bibitem[\citeproctext]{ref-colliander2019comparison}
Colliander, C., \& Ahlgren, P. (2019). Comparison of publication-level approaches to ex-post citation normalization. \emph{Scientometrics}, 120/1: 283--300. DOI: \href{https://doi.org/10.1007/s11192-019-03121-z}{10.1007/s11192-019-03121-z}

\bibitem[\citeproctext]{ref-dickersin2002problems}
Dickersin, K., Scherer, R., Suci, E. S. T., \& Gil-Montero, M. (2002). Problems with indexing and citation of articles with group authorship. \emph{JAMA}, 287/21: 2772--4. DOI: \href{https://doi.org/10.1001/jama.287.21.2772}{10.1001/jama.287.21.2772}

\bibitem[\citeproctext]{ref-donner2016missing}
Donner, P. (2016). Missing citations due to exact reference matching: Analysis of a random sample from {WoS}. Are publications from peripheral countries disadvantaged? \emph{{21st International Conference on Science and Technology Indicators-STI 2016. Book of Proceedings}}, pp. 85--9.

\bibitem[\citeproctext]{ref-donner2017document}
------. (2017). Document type assignment accuracy in the journal citation index data of {Web of Science}. \emph{Scientometrics}, 113/1: 219--36. DOI: \href{https://doi.org/10.1007/s11192-017-2483-y}{10.1007/s11192-017-2483-y}

\bibitem[\citeproctext]{ref-donner2020comparing}
Donner, P., Rimmert, C., \& Van Eck, N. J. (2020). Comparing institutional-level bibliometric research performance indicator values based on different affiliation disambiguation systems. \emph{Quantitative Science Studies}, 1/1: 150--70. DOI: \href{https://doi.org/10.1162/qss_a_00013}{10.1162/qss\_a\_00013}

\bibitem[\citeproctext]{ref-dubin2004most}
Dubin, D. (2004). {The most influential paper Gerard Salton never wrote}. \emph{Library Trends}, 52/4: 748--64.

\bibitem[\citeproctext]{ref-dyachenko2010scientometric}
Dyachenko, E., \& Pislyakov, V. (2010). Scientometric journals in scientometric databases. Microlevel analysis of lost citations. \emph{11th international conference on science and technology indicators. Book of abstracts}, pp. 84--6.

\bibitem[\citeproctext]{ref-efron1994introduction}
Efron, B., \& Tibshirani, R. J. (1993). \emph{An introduction to the bootstrap}. Springer Science+Business Media Dordrecht.

\bibitem[\citeproctext]{ref-franceschini2013novel}
Franceschini, F., Maisano, D., \& Mastrogiacomo, L. (2013). A novel approach for estimating the omitted-citation rate of bibliometric databases with an application to the field of bibliometrics. \emph{Journal of the American Society for Information Science and Technology}, 64/10: 2149--56. DOI: \href{https://doi.org/10.1002/asi.22898}{10.1002/asi.22898}

\bibitem[\citeproctext]{ref-franceschini2015errors}
------. (2015). Errors in {DOI} indexing by bibliometric databases. \emph{Scientometrics}, 102/3: 2181--6. DOI: \href{https://doi.org/10.1007/s11192-014-1503-4}{10.1007/s11192-014-1503-4}

\bibitem[\citeproctext]{ref-franceschini2016museum}
------. (2016). The museum of errors/horrors in {Scopus}. \emph{Journal of Informetrics}, 10/1: 174--82. DOI: \href{https://doi.org/10.1016/j.joi.2015.11.006}{10.1016/j.joi.2015.11.006}

\bibitem[\citeproctext]{ref-garcia2010accuracy}
García-Pérez, M. A. (2010). Accuracy and completeness of publication and citation records in the {Web of Science}, {PsycINFO}, and {Google Scholar}: A case study for the computation of h indices in psychology. \emph{Journal of the American society for information science and technology}, 61/10: 2070--85. DOI: \href{https://doi.org/10.1002/asi.21372}{10.1002/asi.21372}

\bibitem[\citeproctext]{ref-garcia2011strange}
------. (2011). Strange attractors in the {Web of Science} database. \emph{Journal of Informetrics}, 5/1: 214--8. DOI: \href{https://doi.org/10.1016/j.joi.2010.07.006}{10.1016/j.joi.2010.07.006}

\bibitem[\citeproctext]{ref-garcia2015online}
------. (2015). Online supplemental information: A sizeable black hole for citations. \emph{Scientometrics}, 102/2: 1655--9. DOI: \href{https://doi.org/10.1007/s11192-014-1348-x}{10.1007/s11192-014-1348-x}

\bibitem[\citeproctext]{ref-garfield1983quality}
Garfield, E. (1983). Quality control at ISI: A piece of your mind can help us on our quest error-free bibliographic information. \emph{Current Contents}, 19: 5--12.

\bibitem[\citeproctext]{ref-haunschild2022scores}
Haunschild, R., Daniels, A. D., \& Bornmann, L. (2022). Scores of a specific field-normalized indicator calculated with different approaches of field-categorization: Are the scores different or similar? \emph{Journal of Informetrics}, 16/1: 101241. DOI: \href{https://doi.org/10.1016/j.joi.2021.101241}{10.1016/j.joi.2021.101241}

\bibitem[\citeproctext]{ref-haunschild2018algorithmically}
Haunschild, R., Schier, H., Marx, W., \& Bornmann, L. (2018). Algorithmically generated subject categories based on citation relations: An empirical micro study using papers on overall water splitting. \emph{Journal of Informetrics}, 12/2: 436--47. DOI: \href{https://doi.org/10.1016/j.joi.2018.03.004}{10.1016/j.joi.2018.03.004}

\bibitem[\citeproctext]{ref-held2021challenges}
Held, M., Laudel, G., \& Gläser, J. (2021). Challenges to the validity of topic reconstruction. \emph{Scientometrics}. DOI: \href{https://doi.org/10.1007/s11192-021-03920-3}{10.1007/s11192-021-03920-3}

\bibitem[\citeproctext]{ref-hicks2012performance}
Hicks, D. (2012). Performance-based university research funding systems. \emph{Research policy}, 41/2: 251--61. DOI: \href{https://doi.org/10.1016/j.respol.2011.09.007}{10.1016/j.respol.2011.09.007}

\bibitem[\citeproctext]{ref-hicks2015LM}
Hicks, D., Wouters, P., Waltman, L., de Rijcke, S., \& Rafols, I. (2015). Bibliometrics: The {Leiden Manifesto} for research metrics. \emph{Nature}, 520/7548: 429--31. DOI: \href{https://doi.org/10.1038/520429a}{10.1038/520429a}

\bibitem[\citeproctext]{ref-huang2020comparison}
Huang, C.-K., Neylon, C., Brookes-Kenworthy, C., Hosking, R., Montgomery, L., Wilson, K., \& Ozaygen, A. (2020). Comparison of bibliographic data sources: Implications for the robustness of university rankings. \emph{Quantitative Science Studies}, 1/2: 445--78. DOI: \href{https://doi.org/10.1162/qss_a_00031}{10.1162/qss\_a\_00031}

\bibitem[\citeproctext]{ref-jcgm2008evaluation}
JCGM. (2008). \emph{Evaluation of measurement data -- guide to the expression of uncertainty in measurement} ( No. JCGM 100). Joint Committee for Guides in Metrology.

\bibitem[\citeproctext]{ref-jergas2015quotation}
Jergas, H., \& Baethge, C. (2015). Quotation accuracy in medical journal articles---a systematic review and meta-analysis. \emph{PeerJ}, 3: e1364. DOI: \href{https://doi.org/10.7717/peerj.1364}{10.7717/peerj.1364}

\bibitem[\citeproctext]{ref-klavans2017type}
Klavans, R., \& Boyack, K. W. (2017). Which type of citation analysis generates the most accurate taxonomy of scientific and technical knowledge? \emph{Journal of the Association for Information Science and Technology}, 68/4: 984--98. DOI: \href{https://doi.org/10.1002/asi.23734}{10.1002/asi.23734}

\bibitem[\citeproctext]{ref-krauskopf2019missing}
Krauskopf, E. (2019). {Missing documents in Scopus: the case of the journal Enfermeria Nefrologica}. \emph{Scientometrics}, 119/1: 543--7. DOI: \href{https://doi.org/10.1007/s11192-019-03040-z}{10.1007/s11192-019-03040-z}

\bibitem[\citeproctext]{ref-krauskopf2023inconsistency}
Krauskopf, E., \& Salgado, M. (2023). {Inconsistency in the registration of the Digital Object Identifier (DOI) of articles on Web of Science and Scopus}. \emph{Investigaci{ó}n Bibliotecol{ó}gica: archivonom{í}a, bibliotecolog{í}a e informaci{ó}n}, 37/96: 129--44. DOI: \href{https://doi.org/10.22201/iibi.24488321xe.2023.96.58784}{10.22201/iibi.24488321xe.2023.96.58784}

\bibitem[\citeproctext]{ref-leydesdorff2016professional}
Leydesdorff, L., Wouters, P., \& Bornmann, L. (2016). Professional and citizen bibliometrics: Complementarities and ambivalences in the development and use of indicators---a state-of-the-art report. \emph{Scientometrics}, 109/3: 2129--50. DOI: \href{https://doi.org/10.1007/s11192-016-2150-8}{10.1007/s11192-016-2150-8}

\bibitem[\citeproctext]{ref-liu2020accuracy}
Liu, W. (2020). Accuracy of funding information in {Scopus}: A comparative case study. \emph{Scientometrics}, 124/1: 803--11. DOI: \href{https://doi.org/10.1007/s11192-020-03458-w}{10.1007/s11192-020-03458-w}

\bibitem[\citeproctext]{ref-liu2021same}
Liu, W., Huang, M., \& Wang, H. (2021). Same journal but different numbers of published records indexed in {Scopus and Web of Science Core Collection}: Causes, consequences, and solutions. \emph{Scientometrics}, 126/5: 4541--50. DOI: \href{https://doi.org/10.1007/s11192-021-03934-x}{10.1007/s11192-021-03934-x}

\bibitem[\citeproctext]{ref-moed2005citation}
Moed, H. F. (2005). \emph{Citation analysis in research evaluation}. Springer Science \& Business Media. DOI: \href{https://doi.org/10.1007/1-4020-3714-7}{10.1007/1-4020-3714-7}

\bibitem[\citeproctext]{ref-moed1989possible}
Moed, H. F., \& Vriens, M. (1989). Possible inaccuracies occurring in citation analysis. \emph{Journal of Information Science}, 15/2: 95--107. DOI: \href{https://doi.org/10.1177/016555158901500205}{10.1177/016555158901500205}

\bibitem[\citeproctext]{ref-olensky2015data}
Olensky, M. (2015). \emph{Data accuracy in bibliometric data sources and its impact on citation matching} (PhD thesis). Humboldt-Universit{ä}t zu Berlin, Philosophische Fakult{ä}t I. DOI: \href{https://doi.org/10.18452/17122}{10.18452/17122}

\bibitem[\citeproctext]{ref-olensky2016evaluation}
Olensky, M., Schmidt, M., \& Van Eck, N. J. (2016). Evaluation of the citation matching algorithms of {CWTS} and {iFQ} in comparison to the {Web of Science}. \emph{Journal of the Association for Information Science and Technology}, 67/10: 2550--64. DOI: \href{https://doi.org/10.1002/asi.23590}{10.1002/asi.23590}

\bibitem[\citeproctext]{ref-pranckute2021web}
Pranckutė, R. (2021). {Web of Science (WoS) and Scopus: The titans of bibliographic information in today's academic world}. \emph{Publications}, 9/1: 12. DOI: \href{https://doi.org/10.3390/publications9010012}{10.3390/publications9010012}

\bibitem[\citeproctext]{ref-priem2022openalex}
Priem, J., Piwowar, H., \& Orr, R. (2022). {OpenAlex: A fully-open index of scholarly works, authors, venues, institutions, and concepts}. \emph{arXiv preprint arXiv:2205.01833}. DOI: \href{https://doi.org/10.48550/arXiv.2205.01833}{10.48550/arXiv.2205.01833}

\bibitem[\citeproctext]{ref-robinson2023errors}
Robinson-Garcia, N., Torres-Salinas, D., Vargas-Quesada, B., Chinchilla-Rodríguez, Z., \& Gorraiz, J. (2023). Errors of measurement in scientometrics: Identification and calculation of systematic errors. \emph{{Proceedings of ISSI 2023 -- the 19th International Conference of the International Society for Scientometrics and Informetrics, 2}}, pp. 387--93. {International Society for Scientometrics and Informetrics}. DOI: \href{https://doi.org/10.5281/zenodo.8428899}{10.5281/zenodo.8428899}

\bibitem[\citeproctext]{ref-ruiz2015field-normalized}
Ruiz-Castillo, J., \& Waltman, L. (2015). Field-normalized citation impact indicators using algorithmically constructed classification systems of science. \emph{Journal of Informetrics}, 9/1: 102--17. DOI: \href{https://doi.org/10.1016/j.joi.2014.11.010}{10.1016/j.joi.2014.11.010}

\bibitem[\citeproctext]{ref-sauvayre2022misreferencing}
Sauvayre, R. (2022). Misreferencing practice of scientists: Inside researchers' sociological and bibliometric profiles. \emph{Social Epistemology}, 66/6: 719--30. DOI: \href{https://doi.org/10.1080/02691728.2021.2022807}{10.1080/02691728.2021.2022807}

\bibitem[\citeproctext]{ref-schmidt2018fehler}
Schmidt, F. (2018). \emph{{Fehlerabschätzungen bei bibliometrischen Analysen}} (Master's thesis). Technische Hochschule Köln.

\bibitem[\citeproctext]{ref-schneider2013caveats}
Schneider, J. W. (2013). Caveats for using statistical significance tests in research assessments. \emph{Journal of Informetrics}, 7/1: 50--62. DOI: \href{https://doi.org/10.1016/j.joi.2012.08.005}{10.1016/j.joi.2012.08.005}

\bibitem[\citeproctext]{ref-schubert1983statistical}
Schubert, A., \& Glänzel, W. (1983). Statistical reliability of comparisons based on the citation impact of scientific publications. \emph{Scientometrics}, 5/1: 59--73.

\bibitem[\citeproctext]{ref-schubert1988against}
Schubert, A., Glänzel, W., \& Braun, T. (1988). Against absolute methods: Relative scientometric indicators and relational charts as evaluation tools. \emph{Handbook of quantitative studies of science and technology}, pp. 137--76. Elsevier.

\bibitem[\citeproctext]{ref-schulz2016using}
Schulz, J. (2016). Using {Monte Carlo} simulations to assess the impact of author name disambiguation quality on different bibliometric analyses. \emph{Scientometrics}, 107/3: 1283--98. DOI: \href{https://doi.org/10.1007/s11192-016-1892-7}{10.1007/s11192-016-1892-7}

\bibitem[\citeproctext]{ref-shu2019comparing}
Shu, F., Julien, C.-A., Zhang, L., Qiu, J., Zhang, J., \& Larivière, V. (2019). Comparing journal and paper level classifications of science. \emph{Journal of Informetrics}, 13/1: 202--25. DOI: \href{https://doi.org/10.1016/j.joi.2018.12.005}{10.1016/j.joi.2018.12.005}

\bibitem[\citeproctext]{ref-taylor1997introduction}
Taylor, J. R. (1997). \emph{An introduction to error analysis}. University Science Books.

\bibitem[\citeproctext]{ref-tiesinga2021codata}
Tiesinga, E., Mohr, P. J., Newell, D. B., \& Taylor, B. N. (2021). {CODATA} recommended values of the fundamental physical constants: 2018. \emph{Journal of Physical and Chemical Reference Data}, 50/3: 033105. DOI: \href{https://doi.org/10.1063/5.0064853}{10.1063/5.0064853}

\bibitem[\citeproctext]{ref-traag2019louvain}
Traag, V. A., Waltman, L., \& Van Eck, N. J. (2019). From {Louvain to Leiden}: Guaranteeing well-connected communities. \emph{Scientific Reports}, 9/1: 5233. DOI: \href{https://doi.org/10.1038/s41598-019-41695-z}{10.1038/s41598-019-41695-z}

\bibitem[\citeproctext]{ref-tunger2010delphic}
Tunger, D., Haustein, S., Ruppert, L., Luca, G., \& Unterhalt, S. (2010). "The {Delphic Oracle}" - an analysis of potential error sources in bibliographic databases. \emph{{11th International Conference on Science and Technology Indicators. Book of Abstracts}}, pp. 282--3.

\bibitem[\citeproctext]{ref-tuur2016non}
Tüür-Fröhlich, T. (2016). \emph{The non-trivial effects of trivial errors in scientific communication and evaluation}. vwh Verlag Werner H{ü}lsbusch.

\bibitem[\citeproctext]{ref-valderrama2015systematic}
Valderrama-Zurián, J.-C., Aguilar-Moya, R., Melero-Fuentes, D., \& Aleixandre-Benavent, R. (2015). A systematic analysis of duplicate records in {Scopus}. \emph{Journal of Informetrics}, 9/3: 570--6. DOI: \href{https://doi.org/10.1016/j.joi.2015.05.002}{10.1016/j.joi.2015.05.002}

\bibitem[\citeproctext]{ref-vaneck2017accuracy}
Van Eck, N. J., \& Waltman, L. (2017). \href{https://arxiv.org/abs/1906.07011}{Accuracy of citation data in {Web of Science} and {Scopus}}. \emph{{Proceedings of the 16th International Conference of the International Society for Scientometrics and Informetrics}}, pp. 1087--92.

\bibitem[\citeproctext]{ref-wager2008technical}
Wager, E., \& Middleton, P. (2008). Technical editing of research reports in biomedical journals. \emph{Cochrane Database of Systematic Reviews}, 4. DOI: \href{https://doi.org/10.1002/14651858.MR000002.pub3}{10.1002/14651858.MR000002.pub3}

\bibitem[\citeproctext]{ref-waltman2012leiden}
Waltman, L., Calero-Medina, C., Kosten, J., Noyons, E. C., Tijssen, R. J., van Eck, N. J., Van Leeuwen, T. N., et al. (2012). The {Leiden Ranking} 2011/2012: Data collection, indicators, and interpretation. \emph{Journal of the American Society for Information Science and Technology}, 63/12: 2419--32. DOI: \href{https://doi.org/10.1002/asi.22708}{10.1002/asi.22708}

\bibitem[\citeproctext]{ref-waltman2012new}
Waltman, L., \& Van Eck, N. J. (2012). A new methodology for constructing a publication-level classification system of science. \emph{Journal of the American Society for Information Science and Technology}, 63/12: 2378--92. DOI: \href{https://doi.org/10.1002/asi.22748}{10.1002/asi.22748}

\bibitem[\citeproctext]{ref-waltman2019field}
------. (2019). Field normalization of scientometric indicators. \emph{Springer handbook of science and technology indicators}, pp. 281--300. Springer.

\bibitem[\citeproctext]{ref-waltman2011towards}
Waltman, L., Van Eck, N. J., Van Leeuwen, T. N., Visser, M. S., \& Van Raan, A. F. (2011). Towards a new crown indicator: Some theoretical considerations. \emph{Journal of Informetrics}, 5/1: 37--47. DOI: \href{https://doi.org/10.1016/j.joi.2010.08.001}{10.1016/j.joi.2010.08.001}

\bibitem[\citeproctext]{ref-xu2019types}
Xu, S., Hao, L., An, X., Zhai, D., \& Pang, H. (2019). {Types of DOI errors of cited references in Web of Science with a cleaning method}. \emph{Scientometrics}, 120: 1427--37. DOI: \href{https://doi.org/10.1007/s11192-019-03162-4}{10.1007/s11192-019-03162-4}

\bibitem[\citeproctext]{ref-zacharewicz2019performance}
Zacharewicz, T., Lepori, B., Reale, E., \& Jonkers, K. (2019). Performance-based research funding in EU member states---a comparative assessment. \emph{Science and public policy}, 46/1: 105--15. DOI: \href{https://doi.org/10.1093/scipol/scy041}{10.1093/scipol/scy041}

\bibitem[\citeproctext]{ref-zhu2019doi}
Zhu, J., Hu, G., \& Liu, W. (2019). DOI errors and possible solutions for web of science. \emph{Scientometrics}, 118: 709--18. DOI: \href{https://doi.org/10.1007/s11192-018-2980-7}{10.1007/s11192-018-2980-7}

\end{CSLReferences}
\end{document}